\documentclass[12pt,twocolumn]{aastex631}
\usepackage{amsmath}
\usepackage{hyperref}       % load hyperref before cleveref
\usepackage{cleveref}
\usepackage{natbib}
\usepackage{float} 
\usepackage[utf8]{inputenc}
\crefformat{MAPS}{\S#2#1#3}

\usepackage{listings}
\usepackage{xcolor}

\lstdefinestyle{SQLStyle}{
  language=SQL,
  basicstyle=\ttfamily\footnotesize, % smaller font
  keywordstyle=\normalfont,          % remove color and bold from keywords
  commentstyle=\color{gray},
  stringstyle=\normalfont,
  showstringspaces=false,
  breaklines=true,
  breakindent=1em,
  keepspaces=true
}

\shorttitle{Warped \& Hooked: Mapping the Magellanic Clouds in 3D using Red Clump stars}
\shortauthors{Oden et al.}

\begin{document}

% \linenumbers
% https://www.overleaf.com/project/6296e9d0499b8d69ea94c679

\title{Warped \& Hooked: Mapping the Magellanic Clouds in 3D using Red Clump stars}
\author[0009-0000-9037-1697]{Slater J. Oden}
\affiliation{Department of Physics, Montana State University, P.O. Box 173840, Bozeman, MT 59717-3840}
\email{slateroden@montana.edu}

\author[0000-0002-1793-3689]{David L. Nidever}
\affiliation{Department of Physics, Montana State University, P.O. Box 173840, Bozeman, MT 59717-3840}
\affiliation{NSF's National Optical-Infrared Astronomy Research Laboratory, 950 N. Cherry Ave., Tucson, AZ 85719, USA}
% \email{david.nidever@montana.edu}

\author[0000-0002-6553-7082]{Joshua Povick}
\affiliation{Lund Observatory, Division of Astrophysics, Department of Physics, Lund University, Box 118, SE-22100 Lund, Sweden}
% \email{joshua.povick@fysik.lu.se}

\author[0000-0002-8093-7471]{Pol Massana}
\affiliation{CTIO, NOIRLab, Chile}
\affiliation{Department of Physics, Montana State University, P.O. Box 173840, Bozeman, MT 59717-3840}
% \email{polmass@gmail.com}

\author[0000-0002-1594-1466]{Yumi Choi}
\affiliation{NSF's National Optical-Infrared Astronomy Research Laboratory, 950 N. Cherry Ave., Tucson, AZ 85719, USA}
\affiliation{Department of Astronomy, University of California, Berkeley, Berkeley, CA 94720, USA}
% \email{yumi.choi@noirlab.edu}

\author[0000-0001-7827-7825]{Roeland P. van der Marel}
\affiliation{Space Telescope Science Institute, 3700 San Martin Drive, Baltimore, MD 21218, USA}
\affiliation{Center for Astrophysical Sciences, The William H. Miller III Department of Physics \& Astronomy, Johns Hopkins University, Baltimore, MD 21218, USA}
% \email{marel@stsci.edu}

\author[0000-0002-1594-1466]{Maria-Rosa L. Cioni}
\affiliation{Leibniz-Institut f\"ur Astrophysik Potsdam, An der Sternwarte 16, D-14482 Potsdam, Germany}
% \email{mcioni@aip.de}

\author[0000-0002-1594-1466]{Joanna Sakowska}
\affiliation{Instituto de Astrof\'isica de Andaluc\'ia (CSIC), Glorieta de la Astronom\'ia, E-18080 Granada, Spain}
% \email{jsakowska@iaa.csic.es}

\author[0000-0002-7134-8296]{Knut A.~G.~Olsen}
\affiliation{NSF's National Optical-Infrared Astronomy Research Laboratory, 950 N. Cherry Ave., Tucson, AZ 85719, USA}
% \email{knut.olsen@noirlab.edu}

\author[0000-0001-8536-0547]{Lara Cullinane}
\affiliation{Leibniz-Institut f\"ur Astrophysik Potsdam, An der Sternwarte 16, D-14482 Potsdam, Germany}
% \email{lcullinane@aip.de}

%--- DELVE Builders (alphabetical order) ---%

\author[0000-0001-6455-9135]{J. A. Carballo-Bello}
\affiliation{Instituto de Alta Investigaci\'on, Universidad de Tarapac\'a, Casilla 7D, Arica, Chile}
% \email{jcarballo@academicos.uta.cl}

\author[0000-0002-1763-4128]{D.~Crnojevi\'c}
\affiliation{Department of Physics \& Astronomy, University of Tampa, 401 West Kennedy Boulevard, Tampa, FL 33606, USA}
% \email{dcrnojevic@ut.edu}

\author[0000-0001-6957-1627]{P.~S.~Ferguson}
\affiliation{DiRAC Institute and the Department of Astronomy, University of Washington, Seattle, WA, USA}
% \email{pferguso@uw.edu}

\author[0000-0002-9144-7726]{C.~E.~Mart\'inez-V\'azquez}
\affiliation{NSF NOIRLab, 670 N. A\textquotesingle ohoku Place, Hilo, Hawai\`i, 96720, USA}
% \email{clara.martinez@noirlab.edu}

\author[0000-0003-0105-9576]{G.~E.~Medina}
\affiliation{Department of Astronomy and Astrophysics, University of Toronto, 50 St. George Street, Toronto ON, M5S 3H4, Canada}
\affiliation{David A. Dunlap Department of Astronomy \& Astrophysics, University of Toronto, 50 St George Street, Toronto ON M5S 3H4, Canada}
% \email{gustavo.medina@utoronto.ca}

\author[0000-0001-9649-4815]{B.~Mutlu-Pakdil}
\affiliation{Department of Physics and Astronomy, Dartmouth College, Hanover, NH 03755, USA}
% \email{Burcin.Mutlu-Pakdil@dartmouth.edu}

\author[0000-0001-9438-5228]{M. Navabi}
\affiliation{Department of Physics, University of Surrey, Guildford GU2 7XH, UK}
% \email{m.navabi@surrey.ac.uk}

\author[0000-0002-6021-8760]{A.~B.~Pace}
\affiliation{Department of Astronomy, University of Virginia, 530 McCormick Road, Charlottesville, VA 22904, USA}
% \email{pvpace1@gmail.com}

\author[0000-0001-5805-5766]{A.~H.~Riley}
\affiliation{Lund Observatory, Division of Astrophysics, Department of Physics, Lund University, SE-221 00 Lund, Sweden}
% \email{alexander.riley@fysik.lu.se}

\author[0000-0003-1479-3059]{Guy S. Stringfellow}
\affiliation{Center for Astrophysics and Space Astronomy, University of Colorado, 389 UCB, Boulder, CO 80309-0389, USA}
% \email{Guy.Stringfellow@Colorado.EDU}

\author[0000-0003-4341-6172]{A.~K.~Vivas}
\affiliation{Cerro Tololo Inter-American Observatory/NSF NOIRLab, Casilla 603, La Serena, Chile}
% \email{kathy.vivas@noirlab.edu}

\begin{abstract}
The Large and Small Magellanic Clouds (LMC and SMC) are the Milky Way's nearest interacting galaxy pair, offering a unique laboratory for studying tidal effects on galactic disks. Despite extensive survey efforts, the three-dimensional geometry of the Clouds, particularly the putative warp of the LMC, remains poorly constrained due to incompleteness in their crowded centers and the low stellar density of their peripheries, which demand wide-field coverage. Using red-clump (RC) stars as standard candles, corrected for age- and metallicity-dependent population effects with empirically calibrated color-magnitude relations and spatially resolved star-formation histories, we construct the most detailed distance map of the Magellanic System to date. Based on $\sim$2.3 million RC stars from \textit{Gaia} DR3 combined with modern reddening maps, we measure median heliocentric distances of $50.62 \pm 2.32$~kpc for the LMC (to $\sim$23$^{\circ}$) and $60.75 \pm 2.85$~kpc for the SMC (to $\sim$12$^{\circ}$). The maps reveal substructures including the LMC Northern Arm, southern hooks, the Magellanic Bridge, and SMC peripheral over-densities, with refreshed distance estimates. Fitting the LMC disk within $7^{\circ}$ yields a global inclination of $i = 25.32^{\circ} \pm 0.10^{\circ}$ and line-of-nodes position angle of $\theta = 142.34^{\circ} \pm 0.21^{\circ}$. Most strikingly, we find the LMC periphery is warped azimuthally into a U-shaped structure reaching vertical amplitudes of $\sim$7~kpc at radii of $\sim$15~kpc. We interpret this warp as a tidal response to the most recent close passage of the SMC $\sim$300 Myr ago, providing new constraints on the orbital history and dynamical evolution of the Magellanic Clouds.
\end{abstract}


% Section #1
\section{Introduction}
\label{sec:intro}

% Paragraph #1 (rewritten)
The Large and Small Magellanic Clouds (hereafter LMC and SMC) are the Milky Way's (MW) nearest massive satellites, residing at heliocentric distances of $\sim$50~kpc \citep{Pietrzynski2019} and $\sim$60~kpc \citep{deGrijs2015}, respectively. Their proximity allows individual stars to be resolved across both systems, making the Magellanic Clouds (MCs) unrivaled laboratories for studies of stellar evolution, interstellar-medium physics, galaxy interactions, and tidal processes on sub-kiloparsec scales.

% Paragraph #2
%%% ---- Is this paragraph necessary ? --- %%%%
%Together, the Magellanic Clouds (MCs) host several prominent features, including the Magellanic Stream \citep[MS;][]{Putman1998,Nidever2008}, a $200^{\circ}$-long trail of H{\scriptsize I} gas that follows their orbit around the MW and contains a recently discovered small stellar counterpart \citep{Zaritsky2020}. The MS also includes the Leading Arm, a structure that precedes the Clouds in their orbit and extends beyond the Galactic plane \citep{Nidever2010}. Another major feature between the Clouds is the Magellanic Bridge (MB), first identified through its H{\scriptsize I} gas content \citep{Hindman1963}, which connects the two galaxies. Today, the MB is recognized as a vast, diffuse structure composed of gas, young stars \citep{Harris2007,Skowron2014,Noel2015,Mackey2017}, and older to intermediate-age stars \citep{Bagheri2013,Carrera2017,Jacyszyn-Dobrzeniecka2017}.

% Paragraph #2 (new)
The LMC is classified as a barred Magellanic spiral with a well‑defined but perturbed stellar disk \citep{devaucouleurs1972,vandermarel2001b,vandermarel2002}.  
Its disk hosts an off‑centered bar \citep{Zhao2000,Rathore2025}, a dominant single spiral arm \citep{Cioni2000a,Elyoussoufi2019}, and an array of low‑surface‑brightness tidal features \citep[e.g.][]{Mackey2016,Mackey2018,Belokurov2019,GaiaCollaboration2021b,Cullinane2022b}.  
The SMC, in contrast, is an irregular dwarf with a large line‑of‑sight (LOS) depth \citep{Subramanian2012,Scowcroft2016,JacyszynDobrzeniecka2017,Ripepi2017,Almeida2024}, reflecting significant tidal disruption by the LMC.  
Hydrodynamical $N$‑body simulations indicate that a direct collision between the Clouds $\sim$100–300 Myr ago can simultaneously generate the LMC’s off‑centered bar, its tilted disk, and the one‑armed spiral pattern \citep{Zhao2000,Olsen2002,Zaritsky2004,Subramaniam2009a,Besla2012,Diaz2012,Salem2015,Choi2018a,Zivick2019,Wang2022}. Complementary work from \citet{Pardy2016} and \citet{Lucchini2021} further explores how repeated LMC–SMC encounters and ram pressure within the MW halo shape both the stellar and gaseous peripheries.

% Paragraph #3
%Dynamical interactions between the Clouds, as well as potentially with the MW, are the leading mechanism for producing their perturbed features \citep{Besla2007,Diaz2012,Salem2015,Zivick2019}. In particular, \citet{Besla2012}, through an analysis of hydrodynamical simulations, showed that the LMC and SMC are likely on their first in-fall into the MW potential. Their results demonstrate that a direct collision between the Clouds $\sim$100–300~Myr ago can naturally explain many of the LMC's morphological features, such as its off-centered, titled stellar bar and single dominant spiral arm. Additionally, simulations have successfully reproduced extended structures including the MS \citep{Wang2022} and the MB \citep{Besla2012,Zivick2018}.

% Paragraph #3
\citet{Mackey2016, Mackey2018} discovered numerous peripheral substructures beyond the nominal LMC disk ($\gtrsim 7^{\circ}$), including two ``hook-like" features in the south and an extended stellar substructure in the northern periphery stretching $\sim$10~kpc eastward, now commonly referred to as the Northern Arm (NA). The NA feature was studied in detail by \citet{Cullinane2022a}, who demonstrated that it shares a similar geometry and metallicity with the outer LMC disk, suggesting it is composed of perturbed disk material. Using red clump (RC) stars, \citet{elyoussoufi2021} derived aggregate distance estimates for the southeastern and southwestern hooks (referred to as southern substructures 1 and 2 in that paper), as well as for the NA (referred to as northern substructure 1), obtaining distances of 53.9 $\pm$ 0.4~kpc, 55.4 $\pm$ 0.5~kpc, and 57.5 $\pm$ 0.9~kpc, respectively. \citet{Cullinane2022a} also derived field-based aggregate distance estimates to the NA to improve the accuracy of 3D cylindrical velocities relative to the LMC disk plane; however, they did not explicitly present individual field distances, instead noting a total distance difference of $\sim$4.0 $\pm$ 1.4~kpc along the length of the feature.

% Paragraph #4
A distance bimodality in the northeastern SMC was first discovered by \citet{Hatzidimitriou1989} using Horizontal Branch and RC stars, revealing a line-of-sight (LOS) depth as large as 17~kpc. Since then, numerous studies \citep[e.g.,][]{Nidever2013,Ripepi2017,Subramanian2017,Muraveva2018a,James2021,elyoussoufi2021,Almeida2024} have further investigated this phenomenon and its implications for the SMC's structure and interaction history. \citet{Nidever2013} confirmed the bimodality using deep photometry from the MAgellanic Periphery Survey (MAPS), finding evidence for two distinct stellar populations at distances of $\sim$55~kpc and $\sim$67~kpc. More recent work by \citet{elyoussoufi2021} reported a distance of 53.7 $\pm$ 0.8~kpc to the SMC’s eastern periphery. Additionally, the authors found evidence of a bimodal RC feature present across the eastern, northern, and southern regions at varying radii, suggesting that tidal interactions have affected the entire galaxy. Furthermore, \citet{Pieres2017} discovered a stellar overdensity in the northern SMC, referred to as the SMC northern overdensity (SMCNOD), which was found to be indistinguishable in age, metallicity, and distance from nearby SMC field stars. Distance estimates for the SMCNOD range from $\sim$61~kpc based on RR Lyrae stars \citep{Prudil2018} to $\sim$65~kpc from RC stars \citep{elyoussoufi2021}.

% Paragraph #5
Traditional distance tracers (e.g., Cepheids, RR Lyrae, eclipsing binaries, tip of the red giant branch (TRGB) stars, or \textit{Gaia} parallaxes) excel in the inner populated main bodies of the Clouds but suffer either from low number statistics at large radii (e.g., Cepheids, RR Lyrae, eclipsing binaries, TRGB stars) or from large individual uncertainties at Magellanic distances (e.g., \textit{Gaia} parallaxes). Intermediate‑age RC stars ($\sim$1.5–10~Gyr) offer a compelling alternative: they are numerous across both the dense inner disks and the diffuse outskirts, and their luminosity function (LF) is intrinsically narrow because core‑helium‑burning stars ignite with nearly fixed core masses \citep[$\sim$0.45~$M_{\odot}$;][]{Girardi1999}. However, RC absolute magnitudes depend systematically on age and metallicity, so population effects—and spatially variable extinction—must be modeled to unlock their full potential \citep{Girardi2016,Ruiz-Dern2018}.

% Paragraph #6
% Early RC‑based MC distances assumed constant intrinsic colors and magnitudes \citep[e.g.,][]{Paczynski1998,Olsen2002}, yielding valuable but spatially incomplete views that were biased in regions of steep age–metallicity gradients. Subsequent work incorporated population corrections \citep{Choi2018a,Saroon2022}, yet still lacked contiguous coverage of the extreme peripheries. As a result, the global 3D structure -- especially the low-surface brightness diffuse peripheries -- has remained only partially constrained.

% Paragraph #6 (rewritten)
Early RC-based distance estimates to the MCs assumed constant intrinsic colors and magnitudes \citep[e.g.,][]{Paczynski1998,Olsen2002}, yielding valuable but spatially incomplete views that were often biased in regions with strong age–metallicity gradients. More recent studies introduced population corrections \citep[e.g.,][]{Choi2018a,Saroon2022}, providing improved distance estimates and the first indications that the LMC outer disk is significantly warped. However, these efforts still lacked contiguous coverage of the extreme peripheries, leaving the global three-dimensional structure—particularly the diffuse, low-surface-brightness outskirts where tidal signatures are expected to be strongest—only partially constrained.

In this study, we overcome these limitations by leveraging \textit{Gaia}~DR3 astrometry and photometry for $\sim$2.3~million Magellanic red clump (RC) stars to map distances and morphology across both Clouds. We demonstrate that \textit{Gaia}’s flux-limited completeness exceeds 90\% for our RC sample, and that crowding-induced incompleteness in the central regions of the Clouds does not bias our RC selection. Extinction corrections are applied using the OGLE-IV reddening map \citep{Skowron2021} and the recalibrated \citet{Schlegel1998} map. We derive \textit{Gaia} G-band absolute magnitudes for our RC sample with a color–absolute magnitude relation calibrated on local Milky Way RC stars \citep{Ruiz-Dern2018}. To account for age-dependent population effects, we derive spatially resolved star formation histories (SFHs) from deep DECam Local Volume Exploration survey of the Magellanic Clouds (DELVE-MC; \citealt{Nidever2025a}), enabling precise corrections for age-driven RC luminosity variations. With these tools, we construct the first contiguous map of the median distances across the Magellanic System, extending to $\sim$23$^{\circ}$ from the LMC and $\sim$12$^{\circ}$ from the SMC. Our results confirm and extend previous detections of the LMC warp \citep{Balbinot2015,Choi2018a,Saroon2022}, showing that it is a fully azimuthal, U-shaped distortion reaching amplitudes of $\sim$7~kpc at projected radii of $\sim$15~kpc. We argue that this globally coherent warp was likely induced by the most recent close interaction between the LMC and SMC about 300 Myr ago, providing direct evidence of their tidal coupling.

This paper is organized as follows. Section~\ref{sec:data} summarizes the data used in our analysis. Section~\ref{sec:methods} outlines the methodology, including extinction corrections, RC selection, crowding and completeness considerations, population-effect modeling, and distance determinations. In Section~\ref{sec:results} we present our main results, including the MC distance map, the inclination and position angle of the LMC disk as a function of radius, and the characterization of the LMC peripheral warp. Section~\ref{sec:discussion} examines systematic uncertainties, compares our results with other distance tracers, and discusses the implications of our findings for the orbital history of the LMC–SMC system. Finally, Section~\ref{sec:conclusion} summarizes our main conclusions.

%In Section~\ref{sec:discussion}, we discuss potential formation mechanisms for the observed periphery features, including a detailed analysis of the LMC’s azimuthal warp. 

% Section #2
\section{Data}
\label{sec:data}

\subsection{Gaia}

% Paragraph #1
The dataset for this study is sourced from the European Space Agency (ESA)\footnote{\url{https://www.esa.int/}} \textit{Gaia} space telescope \citep{GaiaCollaboration2016} Data Release 3 \citep[DR3;][]{GaiaCollaboration2021a}\footnote{\url{https://www.cosmos.esa.int/web/gaia/data-release-3}}, accessible via the \textit{Gaia} archive\footnote{\url{https://gea.esac.esa.int/archive/}}. \textit{Gaia} provides high-precision astrometry and photometry for over 1.5 billion stars in the Local Group. This extensive dataset offers the most precise astrometric measurements to date, enabling efficient separation of Magellanic members from foreground MW stars.

% Paragraph #2
We utilize \textit{Gaia} photometry across all three bands: the broad G-band ($\lambda_{G}$ $\approx$ 330–1050 nm), the blue BP-band ($\lambda_{BP}$ $\approx$ 330–680 nm), and the red RP-band ($\lambda_{RP}$ $\approx$ 640–1050 nm). While \textit{Gaia's} optical bands are more affected by interstellar dust extinction, their extensive wavelength coverage is highly beneficial for improving astrometric solutions \citep{Lindegren2021}. To ensure the reliability of the astrometric measurements—including stellar positions and proper motions (PMs)—we impose selection criteria on the \texttt{re-normalized unit weight error} (\texttt{ruwe}) and \texttt{astrometric\_excess\_noise} provided by \textit{Gaia}, filtering out stars with problematic astrometric solutions. The Appendix provides details on the ADQL query used to extract the data.

% Paragraph #3
Our selection criteria were designed to be deep enough to adequately sample the RC population in the MCs (the median \textit{Gaia} G-band magnitude of the RC for the MCs is $\sim$ 18.9 mag), while also capturing the brighter, more homogeneous red giant branch (RGB) population. We centered the query on $(\alpha,\delta)$ = (81.28$^{\circ}$,--69.78$^{\circ}$) with a radial extent of 30$^{\circ}$ and limited our selection to stars with magnitudes in the range 17 $<$ $G$ $<$ 20.25 and colors in the range 0.0 $<$ $(BP - RP)$ $<$ 4.75. Quality filters excluded sources with \texttt{astrometric\_excess\_noise} $>$ 2, \texttt{ruwe} $>$ 1.4, or null parallax measurements. These constraints yielded an initial catalog of approximately 43 million stars.

\begin{figure*}
\begin{center}
\includegraphics[width=1.0\textwidth]{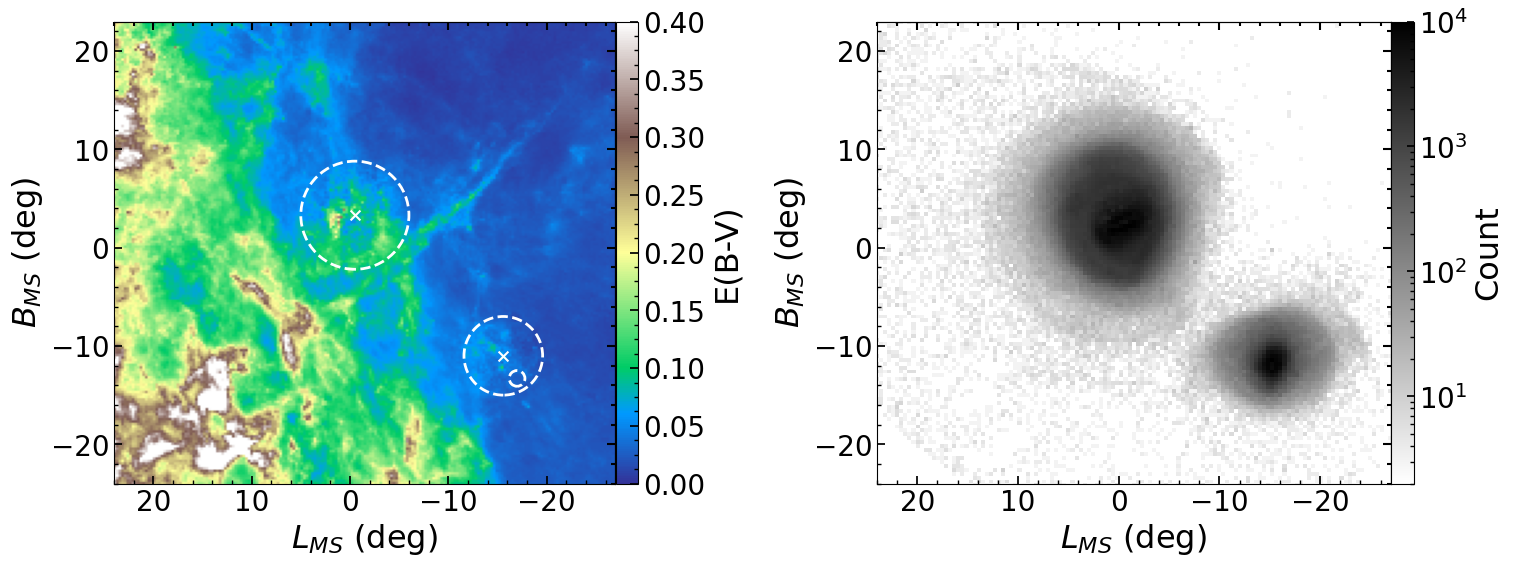}
\caption{\textit{\textbf{Left:}} Combined reddening map of the Magellanic Clouds (MCs) plotted in Magellanic Stream (MS) Coordinates \citep{Nidever2008} using a combination of the OGLE-IV map (inner LMC \& SMC denoted by the dashed white circles) and the recalibrated SFD map in their outer peripheries. The small circle inside the SMC inner region denotes the 47 Tuc region where we adopt SFD values. This reddening map is available publicly on GitHub \hyperlink{https://github.com/slateroden/XMCreddeningmap}{here}. \textit{\textbf{Right:}} 2D density map of the final derived RC sample of the MCs.} 
\label{fig:Extinction_Density_MC}
\end{center}
\end{figure*}

% Paragraph #4
We refine the initial sample by applying a parallax cut of \(\varpi / \sigma_{\varpi} < 4\), which removes nearby MW stars while preserving the intrinsically large distance uncertainties of Magellanic members. We then performed four successive global iterative PM cuts, first selecting stars within a 6~mas~yr\(^{-1}\) radius around the median PM of the sample, followed by progressively stricter cuts with radii of 4~mas~yr\(^{-1}\), 2.5~mas~yr\(^{-1}\) and 1.5~mas~yr\(^{-1}\), re-centering each successive cut on the new median PM of each sample. The final 1.5~mas~yr\(^{-1}\) cut was chosen to sufficiently sample the MC PM distribution, while limiting MW contamination. After these steps, the sample was reduced to \(\sim 7.7\) million stars.

We also obtained a supplementary \textit{Gaia} query centered on $(\alpha,\delta)=(13.10^{\circ},-72.82^{\circ})$ with a radial extent of $15^{\circ}$ to extend our coverage into the SMC periphery. The selection was limited to the same magnitude and color ranges as the primary sample, and sources with problematic astrometric solutions or null parallax measurements were excluded. Identical parallax and global PM cuts were applied, except for the final PM selection, which adopted a radius of 1.2~mas~yr$^{-1}$ to remove obvious MW halo contamination in PM space. After these quality cuts, the sample contained $\sim$1.2~million stars. Merging this catalog with our primary selection and retaining only unique sources resulted in a final combined sample of $\sim$7.8~million stars.

\subsection{DELVE-MC}

Complementary photometry from the DELVE-MC component of the DECam Local Volume Exploration Survey \citep[DELVE;][]{DrlicaWagner2021} is used for SFH derivation to apply second order age-luminosity distance corrections. DELVE-MC provides $\sim$1{,}075 deg$^{2}$ of contiguous $g,r,i$ imaging centered on the MCs, combining new and archival DECam observations to achieve uniform wide-area coverage and depths of $g/r/i \approx 24.8/24.5/24.2$ mag. This dataset represents the deepest panoramic view of the Clouds to date, extending and refining the earlier SMASH survey \citep{Nidever2017,Nidever2021}. Detailed descriptions of the DELVE-MC data products and reduction pipeline will be presented in \citet{Nidever2025a}.

In this work, we employ DELVE-MC photometry primarily for the accurate derivation of spatially resolved SFHs and for calibrating distance corrections. Its combination of depth and spatial coverage enables robust recovery of the oldest main-sequence turnoffs (oMSTO) and low-surface-brightness periphery populations. For this study, we restrict the catalog to stars within $14^{\circ}$ of the LMC and $12^{\circ}$ of the SMC and apply basic quality cuts following the methodology of \citet{Massana2022}, including $-1.0 < {\tt sharp} < 1.0$, ${\tt chi} < 0.3$, and ${\tt prob} > 0.7$. These filters yield a clean point-source catalog ideally suited for SFH fitting and distance calibration.

% \begin{figure*}
% \begin{center}
% \includegraphics[width=1.0\textwidth]{Figures/Reddening_Map_and_Density_XMCS_circles_Xs_colorbar_2SigmaCut.png}
% \caption{\textit{Left:} Combined reddening map of the Magellanic Clouds (MCs) plotted in Magellanic Stream (MS) Coordinates \citep{Nidever2008} using a combination of the OGLE-IV map (inner LMC \& SMC denoted by the dashed white circles) and the recalibrated SFD map in their outer peripheries. This reddening map is available publicly on \hyperlink{https://github.com/slateroden/XMCreddeningmap}{github}. \textit{Right:} 2D Density map of the final derived RC sample of the MCs.} 
% \label{fig:Extinction_Density_MC}
% \end{center}
% \end{figure*}

% Section #3
\section{Methods}
\label{sec:methods}

In this section we describe our methods for extinction correction (Section~\ref{sec:extinction}); selection of Magellanic RC stars (Section~\ref{sec:RC_selection}); crowding and completeness considerations (Section~\ref{sec:completeness}); identification and mitigation of population-driven systematics (Section~\ref{sec:population_systematics}); calibration of the RC absolute magnitude and initial distance estimates (Section~\ref{sec:abs_mag_calibration}); and second-order distance corrections based on SFHs (Section~\ref{sec:SFH_RC_ages}).

% for distance determinations including extinction correction, modeling the RC distance distribution, 

% Section #3.1
\subsection{Extinction Correction}
\label{sec:extinction}

% Paragraph #1
In order to derive distances for individual RC stars in the MCs, we need to correct for interstellar dust extinction in the observed photometry. The amount of reddening toward the MCs is relatively low and has recently been well constrained in the inner regions of the Clouds using the fourth phase of the Optical Gravitational Lensing Experiment (OGLE-IV) Survey \citep{Udalski2015}. Additionally, the widely used \citet{Schlegel1998} reddening map (hereafter SFD map) has been recalibrated \citep{Schlafly2010,Schlafly2011}, indicating a 14\% decrease in the original $E(B-V)$ SFD values and the adoption of a \citet{Fitzpatrick1999} extinction law (hereafter F99 law). Shown previously \citep[e.g.,][]{Choi2018a} the recalibrated SFD map is reliable in the outer regions of the MCs where dust extinction is low. Therefore, by leveraging both reddening maps -- using OGLE-IV values for the inner LMC+SMC and the recalibrated SFD values for the outer periphery -- we construct a spatially contiguous reddening map that accounts for dust variations across the MCs.

%In Section~\ref{sec:Distance_determination} we discuss additional corrections for residual internal reddening in the Clouds that was not captured by these maps.

%\citet{Schlafly2010} notes that the dust extinction spectrum is relatively stable in regions where $E(B-V)$ $\leq$ 0.5~mag (i.e., all fields used in this analysis) and objects can be dereddened in the optical under a universal extinction law with an accuracy of a few percent.

% Paragraph #2
The left panel of Figure~\ref{fig:Extinction_Density_MC} presents the combined reddening map\footnote{This reddening map is available for public use on \hyperlink{https://github.com/slateroden/XMCreddeningmap}{GitHub}}, plotted in Magellanic Stream (MS) coordinates \citep{Nidever2008}. The inner regions of the LMC ($<$ 5.5$^{\circ}$) and SMC ($<$ 4$^{\circ}$) are from the optical $E(V-I)$ reddening map produced by \citet{Skowron2021} using OGLE-IV. We convert the OGLE-IV $E(V-I)$ values to $E(B-V)$ using the relation $E(B-V) = E(V-I)/1.237$ \citep{Schlafly2011}. The resolution of the OGLE-IV map varies from 1$\arcmin$.7 $\times$ 1$\arcmin$.7 in the central parts of the MCs to approximately 27$\arcmin$ $\times$ 27$\arcmin$ in their outskirts. We adopt SFD reddening values in the OGLE-IV gap in the location of the MW globular cluster 47 Tuc. 

% Paragraph #3
To ensure a smooth transition between the OGLE-IV and SFD reddening maps, we apply the logistic weighting function:

\begin{equation}\label{eq:omega}
w(r) = \frac{1}{1 + \exp\left(\frac{r - r_{\text{circle}}}{\omega}\right)}
\end{equation}

\noindent where $r$ is the angular distance from the center of the LMC or SMC, $r_{\text{circle}}$ defines the boundary at which the transition occurs (5.5$^{\circ}$ from the LMC and 4$^{\circ}$ from the SMC), and $\omega$ controls the transition width (i.e., a smaller $\omega$ transitions quickly between maps where a larger $\omega$ transitions more smoothly). For both galaxies we adopt $\omega$ = 2$^{\circ}$. This value is somewhat arbitrary as both reddening maps agree well with each other at their boundaries \citep{Skowron2021} and the final choice of $\omega$ does not effect our final results. Equation~\ref{eq:omega} is applied both across the LMC+SMC (larger white circles in the left panel of Figure~\ref{fig:Extinction_Density_MC}) and within the 47 Tuc gap region (small white circle in Figure~\ref{fig:Extinction_Density_MC}). The final stitched $E(B-V)$ map is then computed as:  

\begin{equation}\label{eq:final_reddening}
\begin{split}
E(B-V) &= w(r) \cdot E(B-V)_{\text{OGLE-IV}} \\
       &\quad + \, (1 - w(r)) \cdot E(B-V)_{\text{SFD}}
\end{split}
\end{equation}

% Paragraph #4
By implementing this weighting scheme, we construct a unified reddening map that preserves the high-resolution structure of the inner Clouds while seamlessly incorporating the corrected SFD map in the periphery. We calculate a median reddening value of $\sim$0.07~mag across the MCs, consistent with previous works (\citealp[]{Haschke2011,Choi2018a,Gorski2020,Skowron2021}).

% Given that the median absolute deviation between OGLE-IV and SFD reddening values is approximately 0.01 mag—comparable to the \textit{Gaia} photometric uncertainty at $G \sim 18$–$19$ mag—this approach provides a robust method for obtaining a unified reddening map across the entire Magellanic system.

% Paragraph #5
To correct for extinction in the \textit{Gaia} passbands, we adopt the F99 \citep{Fitzpatrick1999} extinction law, which parameterize interstellar extinction as a function of the total-to-selective extinction ratio, $R_V$. We assume a standard value of $R_V = 3.1$, appropriate for diffuse MW dust, and derive extinction coefficients for the \textit{Gaia} $G$, $BP$, and $RP$ bands using the \texttt{dust\_extinction} Python package. Specifically, we compute the extinction curve at the effective wavelength of each band, scale the resulting $A_\lambda/A_V$ ratios by $A_V = R_V \times E(B-V)$, and thereby obtain absolute extinctions. This approach follows the recalibrated extinction curves of \citet{Schlafly2011}, ensuring consistency with prior \textit{Gaia}-based studies.  

\begin{subequations}
\label{eq:extinction_laws}
\begin{align}
A_G   &= 0.73165 \cdot R_V \cdot E(B-V), \label{eq:Ag} \\
A_{BP} &= 1.02557 \cdot R_V \cdot E(B-V), \label{eq:Abp} \\
A_{RP} &= 0.55644 \cdot R_V \cdot E(B-V). \label{eq:Arp}
\end{align}
\end{subequations}

\begin{figure}
\includegraphics[width=0.5\textwidth]{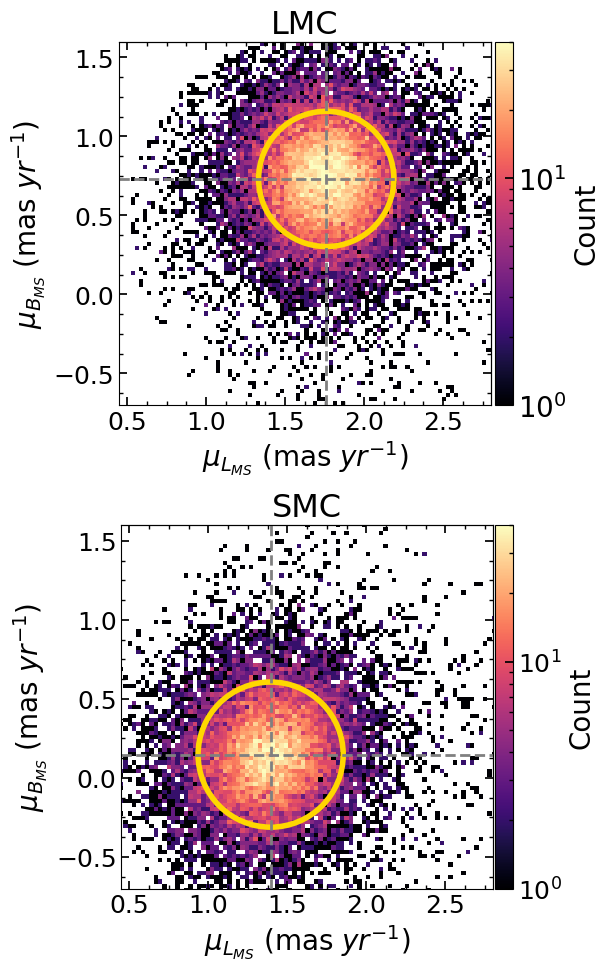}
\caption{Example localized proper motion (PM) cuts in the nearest neighbor spatial search for the LMC (top) and SMC (bottom). The gold circles define the local 2$\sigma$ mas yr$^{-1}$ PM cut centered on the median PM of the local nearest neighbor searches.}
\label{fig:PM_cut}
\end{figure}

% Section #3.2
\subsection{Selecting the Magellanic RC}
\label{sec:RC_selection}

% \begin{figure*}
% \begin{center}
% \includegraphics[width=0.75\textwidth]{Figures/4_Panel_LMC_SMC_CMD_LF_color_FITTING.png}
% \caption{Zoomed in CMDs of representative grid cells showing our selection procedure of Magellanic RC stars. The solid (dashed) red polygons are the limits of the best fit mean gaussian(s) in the LMC (SMC). The top marginal histograms show the distribution in $(BP - RP)_{o}$ color, while the right marginal histograms show the distribution in \textit{Gaia} $G_{o}$-band apparent magnitude. The solid black curves show the total combined fits. The blue (red) dashed lines show the best fit color Gaussian(s) components over the RC (RGB). The solid grey histograms indicate our selection of individual RC star candidates in magnitude and color. Only stars that pass both selections in magnitude and color are adopted as Magellanic RC star candidates. \textit{Top Left:} Example cell in the main disk of the LMC. \textit{Top Right:} Example cell in the main body of the SMC. \textit{Bottom Left:} Example cell in the dense inner bar of the LMC. \textit{Bottom Right:} Example cell in the eastern SMC presenting the bi-modal RC population.}
% \label{fig:CMD_cut}
% \end{center}
% \end{figure*}

\begin{figure*}
\begin{center}
\includegraphics[width=0.75\textwidth]{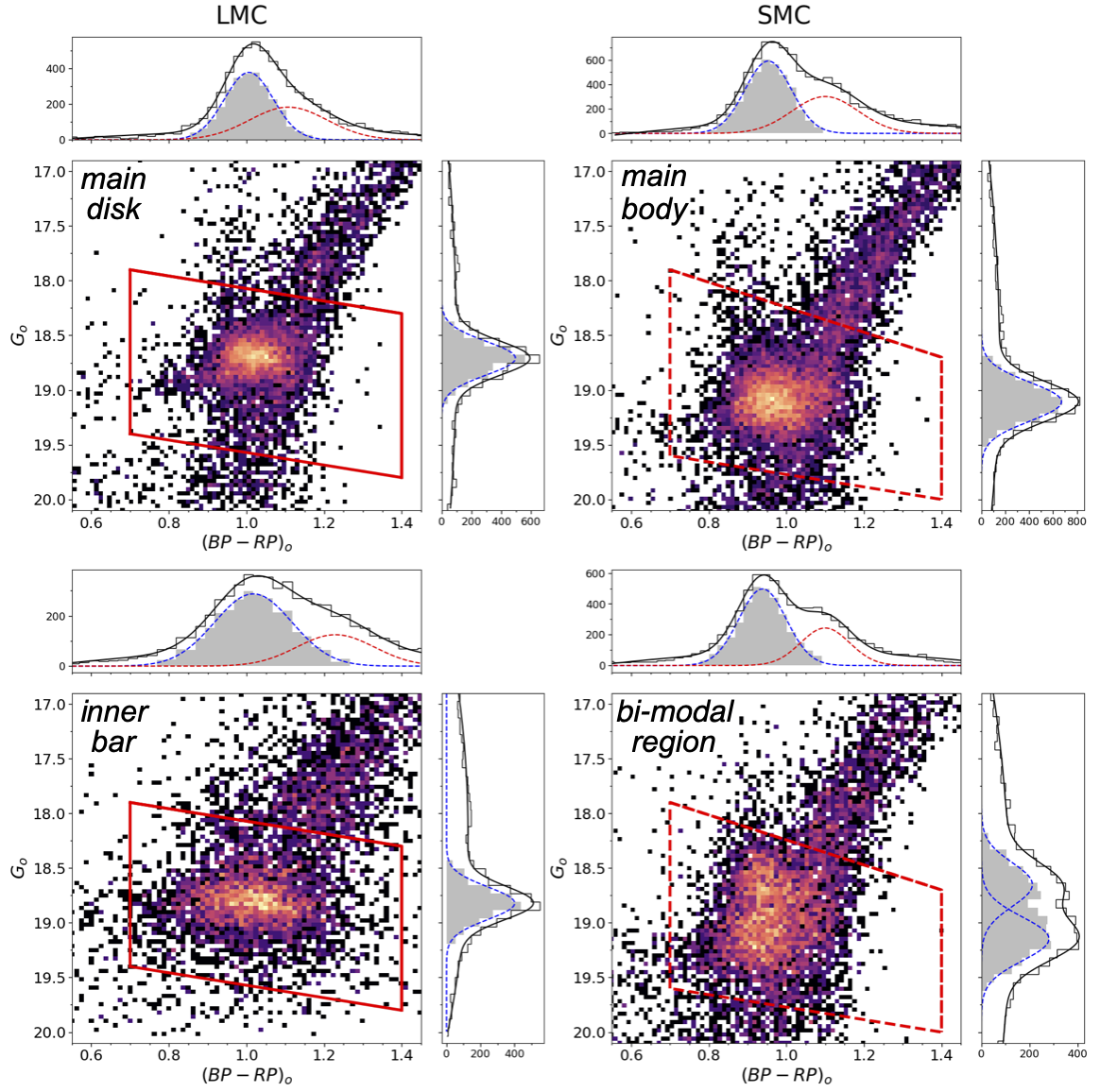}
\caption{Zoomed in CMDs of representative grid cells showing the selection procedure of Magellanic RC stars. The solid (dashed) red polygons are the limits of the best fit mean gaussian(s) in the LMC (SMC). The top marginal histograms show the distribution in $(BP - RP)_{o}$ color, while the right marginal histograms show the distribution in \textit{Gaia} $G_{o}$-band apparent magnitude. The solid black curves show the total combined fits. The blue (red) dashed lines show the best fit color Gaussian(s) components over the RC (RGB). The solid grey histograms indicate our selection of individual RC star candidates in magnitude and color. Only stars that pass both selections in magnitude and color are adopted as Magellanic RC star candidates. \textit{\textbf{Top Left:}} Example cell in the main disk of the LMC. \textit{\textbf{Top Right:}} Example cell in the main body of the SMC. \textit{\textbf{Bottom Left:}} Example cell in the dense inner bar of the LMC. \textit{\textbf{Bottom Right:}} Example cell in the eastern SMC presenting the bi-modal RC population.}
\label{fig:CMD_cut}
\end{center}
\end{figure*}

% Paragraph #1
The RC is one of the most distinct and well-defined features in the CMD of intermediate-age to old stellar populations ($\sim$1.5--10~Gyr), making it ideal for straightforward selection. Its exact position in an extinction-corrected (de-reddened) CMD is primarily determined by the galaxy’s SFH, age–metallicity relation (AMR), and line-of-sight distance. Because these properties vary across the MCs, the main SMC RC population appears slightly bluer and fainter than the main LMC RC population, reflecting its lower metallicity and greater average distance. There also exists metallicity and distance gradients throughout the LMC and SMC, although to a lesser extent, effecting the location of RC stars in the CMD within the Clouds themselves. Therefore, we must account for this spatial variation when selecting RC stars in the MCs.

% Paragraph #2
After correcting for extinction described in Section~\ref{sec:extinction}, we first apply a broad CMD cut targeting Magellanic red giants (primarily RGB and RC stars). This photometric cut, along with our initial parallax filter and global iterative PM cuts (described in Section~\ref{sec:data}), significantly reduces MW contamination, which is most problematic in the outer periphery of the Clouds. To remove residual MW contaminants, we apply a local PM selection (see Figure~\ref{fig:PM_cut}). The PM distribution varies across the MCs—particularly in the LMC, which exhibits strong inner rotation \citep{Choi2022}. The SMC’s mean PM also differs from that of the LMC. As a result, a single global PM cut cannot simultaneously encompass both galaxies while effectively minimizing MW contamination.

% Paragraph #3
We separate the LMC and SMC spatially using a 13$^{\circ}$ circular cut centered on $(L_{\mathrm{MS}}, B_{\mathrm{MS}}) = (-19^{\circ}, -12.5^{\circ})$, chosen to bisect the region of lowest stellar number density between the Clouds. Next, we construct a grid with 0.3$^{\circ}$ resolution across each galaxy, selected to balance statistical robustness with spatial resolution. In particular, we adopt a slightly coarser grid to ensure sufficient number counts per bin in the periphery. Using \texttt{scipy.spatial.KDTree}, we perform nearest-neighbor searches to identify giants closest to each grid point, starting with a 0.3$^{\circ}$ radius—appropriate for the dense inner regions. If fewer than 500 stars are found, we incrementally increase the radius in 0.1$^{\circ}$ steps, up to a maximum of 5.0$^{\circ}$, to adequately sample the low-surface-brightness outskirts. These aperture search limits are somewhat arbitrary and were determined through trial and error.

% Paragraph #4
Within each local sample, we apply a circular PM cut of radius $2\sigma_{\mathrm{PM}}$ mas yr$^{-1}$ centered on the sample’s median PM (gold circles in Figure~\ref{fig:PM_cut}), where $\sigma_{\mathrm{PM}}$ is the robust PM standard deviation of the locally selected stars. This local $2\sigma_{\mathrm{PM}}$ PM cut was chosen to balance Magellanic completeness with MW contamination.

% For cells that required a search radius $>0.3^{\circ}$, we invoke \texttt{numpy.unique} on $(\alpha,\delta)$ to retain only unique giants and avoid duplicates. 

% Paragraph #5
To benchmark our adaptive local PM selection, we also tested a single global circular cut with a radius of $2\sigma_{\mathrm{PM,,Total}}$ mas yr$^{-1}$ centered on the overall median PM of the MCs. The resulting globally selected catalog is 22.8\% larger than the final LMC+SMC sample obtained with our local PM method. Because the combined MC PM distribution is non-circular, most of this excess consists of MW contaminants. In contrast, our adaptive local method re-centers the $2\sigma_{\mathrm{PM}}$ mas yr$^{-1}$ aperture on each cell’s median, retaining $>$95\% of high-quality LMC/SMC giants while removing $\gtrsim$20\% more MW contamination than the global cut. Although the conservative local threshold may discard some bona-fide MC members with extreme velocities, it yields a markedly cleaner tracer population. Crucially, the median distances to the Clouds derived from the two PM-cut strategies differ by $<$1\%, demonstrating that our results are robust to the choice of PM-selection method.

% Paragraph #6
To isolate and select the RC, we repeat the localized nearest-neighbor search on the same 0.3$^{\circ}$ grid and model both the LF and the color distribution. For each grid point, we gather giants within a 0.3$^{\circ}$ radius, expanding in 0.1$^{\circ}$ steps (up to a maximum of 3.0$^{\circ}$) until at least 300 stars are obtained. Cells with fewer than 300 stars—typical in the diffuse periphery of the LMC or SMC—are handled using a fallback RC box selection because they lack the necessary number statistics for fitting. For these cells, we adopt fixed selection boxes in color–magnitude space: 18.25 $< G_{0} <$ 19.25 and 0.83 $< (BP - RP)_{0} <$ 1.07 for the LMC, and 18.4 $< G_{0} <$ 19.7 and 0.77 $< (BP - RP)_{0} <$ 1.07 for the SMC. Cells with fewer than 50 stars within the max search radius are excluded from further analysis.

Figure~\ref{fig:CMD_cut} illustrates our RC-selection procedure for representative grid cells. For LMC cells we model each local LF as a single Gaussian plus second-order polynomial, which cleanly isolates the RC peak (Gaussian) from the underlying RGB ``continuum'' (polynomial). For SMC cells we fit both a single-Gaussian and a double-Gaussian model (each with the same polynomial background) and adopt the solution with the lower reduced $\chi^{2}$. The double-Gaussian model is essential for capturing the pronounced RC bimodality observed in the eastern SMC (see the lower-right panel of Figure~\ref{fig:CMD_cut}). We restrict the Gaussian means to lie within the red solid (dashed) polygons for the LMC (SMC) in all grid-cell fits.

% Paragraph #8
The bright- and faint-end magnitude limits of the modeled local RC population varies throughout the Clouds. This is primarily due to changes in RC population effects, residual internal reddening and the geometric LOS depth of the population, as well as a small degree due to photometric errors. To define the magnitude limits of the RC in each cell we begin with the mean, $\mu$, and dispersion, $\sigma$, of the best‐fit Gaussian component(s). Generally, in dense inner cells the intrinsic RC width is well constrained, so a narrower selection is sufficient; in the low‐surface‐brightness outskirts a slightly broader selection is needed to maintain completeness, mainly due to a greater mix of RC stars and the higher LOS depth in the periphery \citep{Alves2000,Subramanian2009b}. We, therefore, scale our magnitude selection, $\Delta m = \sigma_{\rm scale}\,\sigma$, according to the local sample size $N$. Cells with $N = 300$ stars (typical in the periphery) use a $\sigma_{\rm scale}$ = 3.0; while very dense cells with $N > 800$ stars use $\sigma_{\rm scale}$ = 2.5; intermediate densities are interpolated linearly between these values. Our final selection in magnitude is $\mu_{G_{0}} \pm \Delta m  = \mu_{G_{0}} \pm \sigma_{\rm scale}\sigma$ (shown as the gray histograms in Figure~\ref{fig:CMD_cut}).

%Paragraph #9
The color distribution in both galaxies is modeled as a double Gaussian plus second-order polynomial. Because the genuine RC population lies slightly bluer than the RGB sequence, we can partially isolate and exclude RGB contamination in color space. We, therefore, adopt only the bluer Gaussian peak to define the final RC color window (dotted blue curves in Figure~\ref{fig:CMD_cut}). As with the magnitude selection, the color window is scaled according to the local stellar count $N$, giving a final range of $\mu_{(BP-RP)_{0}} \pm \sigma_{\rm scale}\sigma$, where the scaling factor $\sigma_{\rm scale}$ is identical to that used for the magnitude cuts. Only stars that pass both selections in magnitude and color are adopted as ``Magellanic RC star candidates". Through our initial cuts and fine tuned local PM and CMD fitting selections, we obtain $\sim$1.8 million Magellanic RC star candidates for the LMC and $\sim$470 thousand Magellanic RC star candidates for the SMC. The right panel of Figure~\ref{fig:Extinction_Density_MC} shows the 2D density map of our final selected Magellanic RC star candidates plotted in MS coordinates. 

\begin{figure}
% \begin{center}
\includegraphics[width=0.5\textwidth]{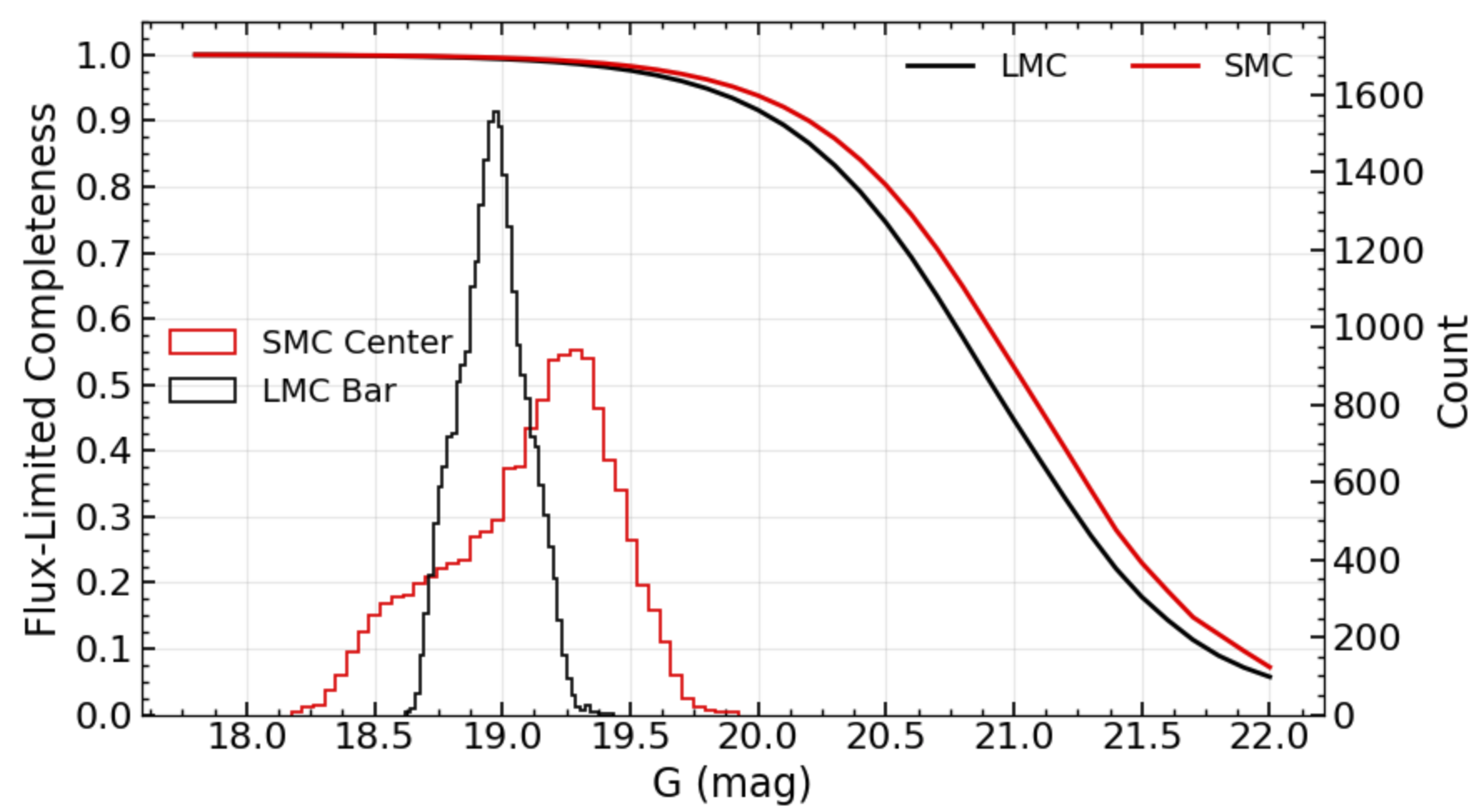}
\caption{\textit{Gaia}'s flux-limited completeness modeled as a sigmoid function, whose shape parameter $\alpha$ is derived from the slope of the faint-end LF falloff. We estimate the 50\% completeness limits for \textit{Gaia} in the LMC and SMC to be approximately $\sim$20.9 and $\sim$21.0 mag, respectively. The non–extinction-corrected $G$-band magnitude distributions for the RC stars selected in the LMC bar (black) and SMC center (red) are also shown, demonstrating that our RC selection remains more than 90\% complete at these magnitudes.} 
\label{fig:completeness}
% \end{center}
\end{figure}

% Paragraph #10
These selection procedures are inevitably subject to some contamination and therefore propagate as systematic uncertainties in distance calculations. \textit{Gaia}'s precision in astrometry and photometry helps mitigate most of the contamination through our CMD and PM cuts. However, given the complex overlap of RC/RGB stars and Magellanic and MW populations, some level of contamination remains unavoidable. Fortunately, distance measurements are not significantly affected by Magellanic RGB contamination \citep{Choi2018a}, as these stars that pass our rigorous CMD selection procedure have similar magnitudes and color as true RC stars. We refer readers to Section~\ref{sec:systematic_errors} for a full discussion of systematic errors. 

%In PM space, the MW halo is relatively evenly dispersed at the $\sim$1\% level, providing an upper limit on the contamination from non-Magellanic members when using a global PM cut. However, given our localized PM cut described above, we expect this contamination to be significantly reduced below 1\%.

% Paragraph #8
%In the CMD, the RGB population overlaps with the RC at the $\sim$10--40\% level \citep{Choi2018a}, leading to higher contamination from RGB stars within the MCs themselves. Modeling the color distribution as a double Gaussian plus second-order polynomial and selecting only the bluer Gaussian peak allows us to remove the bulk of the RGB contaminants. Nevertheless, RGB contamination at the $\sim$5-10\% level remains. Fortunately, distance measurements are not significantly affected by this contamination, as RGB stars that pass our selection procedure have similar magnitudes and color as true RC stars. A full discussion of systematic uncertainties on our results will be presented in Section~\ref{sec:systematic_errors}. 

% Furthermore, we model the resulting contaminant distances as a linear polynomial, effectively eliminating their contribution from our distance maps. This is discussed in detail in Section~\ref{sec:Distance_determination}.

% Section #3.3
\subsection{Crowding \& Completeness}
\label{sec:completeness}

The RC-selection procedures described above assume that the underlying stellar sample is not significantly affected by flux-limited incompleteness across the relevant magnitude range. The standard approach for quantifying survey completeness relies on artificial-star tests (ASTs), which measure the recovery fraction of injected sources as a function of magnitude \citep[e.g.,][]{Rubele2010,Massana2022,Sakowska2024}. While robust, ASTs are computationally expensive and impractical to perform across the full spatial extent of the MCs for this study.

% Instead, we assess \textit{Gaia}’s flux-limited completeness directly from the observed luminosity function (LF). As discussed in greater detail in Section~\ref{sec:SFH_RC_ages}, we model the LF fall-off using a sigmoid recovery function whose shape is determined by the local faint-end LF slope. A sigmoid function has been shown to provide an accurate description of completeness when validated against ASTs \citep[e.g.,][]{Fleming1995,Rejkuba2005,Bonatto2019}, and we have further verified the method against SMASH ASTs. For this test we extract all \textit{Gaia} sources within $3^{\circ}$ of the LMC center, adopted as $(\alpha_{0},\delta_{0})=(80.44^{\circ},-69.87^{\circ})$ 
% \citep{Choi2022} and $2^{\circ}$ from the SMC center, adopted as $(\alpha_{0},\delta_{0})=(13.05^{\circ},-72.83^{\circ})$ \citep{devaucouleurs1972}. The resulting LF's show a smooth exponential decline at faint magnitudes, yielding a 50\% completeness limit of $G \simeq 20.9$~mag for the LMC and $G \simeq 21.0$~mag for the SMC. Figure~\ref{fig:completeness} shows the resulting \textit{Gaia} flux-limited completeness as a function of observed G-band magnitude. This analysis confirms that even at the non-extinction corrected magnitudes of the LMC and SMC RC populations (G$_0 \sim$ 18.5–20.0), \textit{Gaia} is intrinsically $>$98\% complete even in the most crowded regions of the LMC and SMC.

Instead, we estimate \textit{Gaia}’s flux-limited completeness directly from the observed LF. As described in Section~\ref{sec:SFH_RC_ages} for DELVE-MC data, we model the completeness using a sigmoid recovery function whose shape is determined by the local faint-end LF slope. The sigmoid formalism provides an accurate description of completeness when validated against ASTs \citep[e.g.,][]{Fleming1995,Rejkuba2005,Bonatto2019}, and we have further verified our implementation against the Survey of the MAgellanic Stellar History \citep[SMASH;][]{Nidever2017} ASTs. For this test, we extract all \textit{Gaia} sources within $3^{\circ}$ of the LMC center, adopted here as $(\alpha_{0},\delta_{0})=(80.44^{\circ},-69.87^{\circ})$ \citep{Choi2022}, and within $2^{\circ}$ of the SMC center, adopted here as $(\alpha_{0},\delta_{0})=(13.05^{\circ},-72.83^{\circ})$ \citep{devaucouleurs1972}. The resulting LF's exhibit smooth exponential declines toward faint magnitudes, yielding 50\% completeness limits of $G\simeq20.9$~mag for the LMC and $G\simeq21.0$~mag for the SMC. Figure~\ref{fig:completeness} shows the corresponding \textit{Gaia} completeness functions as a function of apparent $G$ magnitude. This analysis confirms that, even at the uncorrected RC magnitudes ($G \!\sim\! 18.5$–$20.0$ mag), \textit{Gaia} remains intrinsically $>$90\% complete at the level of the Magellanic RC.

Despite this high flux-limited completeness at the level of the Magellanic RC, \textit{Gaia}’s on-board flux measurements become increasingly affected by crowding in the high-density cores of the LMC bar and central SMC. In these regions, source losses are dominated by stellar density rather than by the instrument’s flux-detection threshold. Consequently, even with our rigorous RC selection procedure (Section~\ref{sec:RC_selection}), a fraction of stars in the inner LMC and SMC remain undetected due to crowding. To assess whether this crowding-dominated incompleteness biases our RC selection, we estimate the crowding completeness following the method developed by \citet{Rathore2025}, which leverages \textit{Gaia} color-excess measurements.

\begin{figure}
\includegraphics[width=0.5\textwidth]{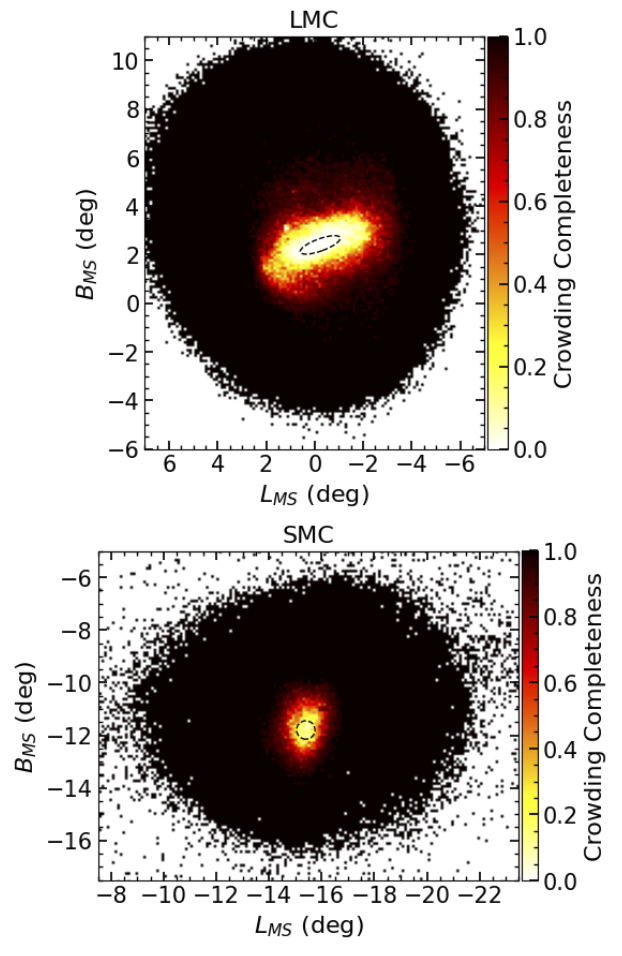}
\caption{\textit{\textbf{Top:}} Crowding-dominated completeness map of RC stars ($G \sim 18.5$--$20.0$ mag) in the LMC. \textit{\textbf{Bottom:}} Same as above, but for the SMC. Values near unity indicate regions where essentially all stars present are detected, whereas values near zero correspond to areas where most stars are missed due to crowding. The dashed ellipse (LMC) and dashed circle (SMC) mark the central regions used to assess potential completeness-driven biases in our RC selection.}
\label{fig:completeness_map}
\end{figure}

For a full description of the method, we refer the reader to \citet{Rathore2025}; here we summarize the key steps. The \textit{Gaia} color excess, $C$, is defined as the ratio of BP$+$RP flux to $G$-band flux. In uncrowded regions, $C$ is slightly greater than unity because the combined BP and RP wavelength coverage exceeds that of the $G$ filter. In crowded fields, contamination from neighboring sources introduces random deviations in $C$, making the scatter in $C$ a direct empirical measure of crowding.

Following \citet{Rathore2025}, we subtract the global color-excess trend in 0.1~mag bins to obtain a corrected color excess, $C^{*}$. We then define an “incomplete” RC sample as the subset of stars with $C^{*}$ $<$ 3$\sigma_{C^{*}}$, where $\sigma_{C^{*}}$ is the robust dispersion of $C^{*}$. This is the sample to which we subsequently apply our completeness corrections, defined as the inverse–completeness weight assigned to each star, yielding the "complete" RC sample.

\begin{table*}
\centering
\caption{Various Center of Mass Values of the Magellanic Clouds}
\label{tab:MC_centers}
\begin{tabular*}{\textwidth}{@{\extracolsep{\fill}} lccc}
\hline\hline
\multicolumn{4}{c}{\textbf{Large Magellanic Cloud (LMC)}} \\
\hline
\textbf{COM Type} & \textbf{Reference} & \textbf{R.A., Decl.} & \textbf{Separation (deg)} \\
\hline
\textbf{Stellar dynamical (original sample)}           
    & \textbf{This work} & (\textbf{80.60$^\circ$}, \textbf{-69.80$^\circ$}) & \textbf{0.00} \\
Stellar dynamical (completeness corrected) 
    & This work & (80.65$^\circ$, $-69.70^\circ$) & 0.10 \\
Stellar dynamical (incomplete)             
    & This work & (80.70$^\circ$, $-70.31^\circ$) & 0.51 \\
Stellar Kinematic     
    & \citet{Choi2022} & (80.44$^\circ$, $-69.27^\circ$) & 0.53 \\
Photometric    
    & \citet{vandermarel2001a} & (81.28$^\circ$, $-69.78^\circ$) & 0.24 \\
H\,\textsc{i} Kinematics    
    & \citet{Kim1998} & (79.35$^\circ$, $-69.03^\circ$) & 0.89 \\
Stellar dynamical (Rathore)
    & \citet{Rathore2025} & (80.27$^\circ$, $-69.65^\circ$) & 0.24 \\
Stellar Kinematic (PMs)
    & \citet{Vijayasree2025} & (80.30$^\circ$, $-69.27^\circ$) & 0.54 \\
\hline
\multicolumn{4}{c}{\textbf{Small Magellanic Cloud (SMC)}} \\
\hline
\textbf{COM Type} & \textbf{Reference} & \textbf{R.A., Decl.} & \textbf{Separation (deg)} \\
\hline
\textbf{Stellar dynamical (original sample)}              
    & \textbf{This work} & (13.21$^\circ$, $-72.97^\circ$) & \textbf{0.00} \\
Stellar dynamical (completeness corrected) 
    & This work & (13.07$^\circ$, $-73.03^\circ$) & 0.07 \\
Stellar dynamical (incomplete)             
    & This work & (13.81$^\circ$, $-73.10^\circ$) & 0.22 \\
Stellar Kinematic     
    & \citet{Piatti2021} & (13.30$^\circ$, $-72.85^\circ$) & 0.12 \\
Photometric    
    & \citet{Mateo1998} & (13.18$^\circ$, $-72.82^\circ$) & 0.15 \\
H\,\textsc{i} Kinematics    
    & \citet{Stanimirovic2004} & (16.25$^\circ$, $-72.42^\circ$) & 1.06 \\
\hline
\end{tabular*}
\end{table*}

Next, we measure the scatter $\sigma_{C^{*}}$ as a function of radius in $0.2^{\circ}$ bins and fit it with a hyperbolic tangent model to determine the intrinsic scatter, $\sigma^{\mathrm{int}}_{C^{*}}$, arising from instrumental and astrophysical affects. We find $\sigma^{\mathrm{int}}_{C^{*}}\simeq0.047$ for the LMC and $\sigma^{\mathrm{int}}_{C^{*}}\simeq0.043$ for the SMC. Subtracting $\sigma^{\mathrm{int}}_{C^{*}}$ in quadrature from the observed $\sigma_{C^{*}}$ yields the crowding-induced scatter, $\sigma^{\mathrm{crowd}}_{C^{*}}$, which we compute in spatial bins of $0.02^{\circ}\!\times\!0.02^{\circ}$ in MS coordinates. Finally, we convert the resulting 2D $\sigma^{\mathrm{crowd}}_{C^{*}}$ map into a crowding completeness map using Equation~5 of \citet{Rathore2025}, calibrated against ASTs from \citet{Olsen2003}. The resulting completeness maps for the RC stars selected in the LMC (top) and SMC (bottom) are shown in Figure~\ref{fig:completeness_map}.

We find that within $2.5^{\circ}$ of the LMC center, the crowding-dominated completeness of our RC sample is $\sim$58\%, slightly lower than the 64\% derived by \citet{Rathore2025}. However, this difference can be expected, as our RC selection includes fainter and redder stars due to our extinction correction prior to our RC selection. For the SMC, the median completeness within $2^{\circ}$ is $\sim$85\%. To our knowledge, no previous study has reported a direct crowding-dominated completeness for RC stars in the SMC core; however, ASTs from the VISTA survey of the Magellanic Clouds (VMC) \citep{Cioni2011} report a completeness within the RC magnitude range 17-18 mag in the $K_{s}$ band of $\sim$91\% $\pm$ 4\% $K_{s}$, $\sim$90\% $\pm$ 14\% in $J$ and $\sim$89\% $\pm$ 14\% in $Y$ for the central SMC tile (tile $4_3$) \citep{Rubele2018}, consistent with our estimate.

Because crowding-dominated completeness primarily affects the inner regions of both Clouds, we isolate stars within an ellipse centered on the LMC bar and within a circular region of radius $0.35^{\circ}$ around the SMC center (see Figure~\ref{fig:completeness_map}). Their corresponding $G$-band magnitude distributions are shown in Figure~\ref{fig:completeness}. To test for potential biases in our RC selection as a function of magnitude, we plot the crowding-dominated completeness against apparent $G$-band magnitude for these inner subsamples. Within the narrow magnitude ranges of our selected RC samples ($\sim$1 mag for the LMC and $\sim$1.5 mag for the SMC), we find no significant correlation between completeness and $G$ magnitude, indicating that our RC selection (Section~\ref{sec:RC_selection}) is not systematically biased toward brighter stars due to crowding. In other words, completeness does not significantly vary with magnitude within these inner RC samples, and therefore our RC selection and distance determinations are not biased toward the brighter portion of the RC.

The crowding and completeness analyses presented here demonstrate that \textit{Gaia} remains highly complete ($>$90\%) across the RC magnitude range, except within the innermost $\sim$$2^{\circ}$ of the LMC and SMC, where crowding marginally reduces completeness. Crucially, we find that this incompleteness does not introduce measurable magnitude–dependent biases in our RC selection. To further verify this, and to derive updated centers for the Clouds, we compute dynamical centers using three samples: the original RC selection, the incomplete RC sample, and the completeness-corrected sample.

% Throughout the remainder of our analysis, stars in the “complete’’ sample are assigned an inverse–completeness weight. In the next section, we test the dynamical centers derived from our original, completeness–corrected, and incomplete RC samples and show that completeness corrections have only a negligible effect on the inferred center of mass. This further verifies that crowding does not significantly bias our RC selection.

% \begin{figure*}
% \begin{center}
% \includegraphics[width=1.0\textwidth]{Figures/2_panel_MC_Centers.png}
% \caption{Center-of-mass estimates for the LMC (left) and SMC (right). The centers derived from our original RC samples are shown as red open circles. Completeness–corrected and incomplete samples are indicated by black “C” markers and dark blue triangles, respectively. Literature estimates are over plotted for comparison: stellar kinematic centers (blue “K”) from \citet{Choi2022} (LMC) and \citet{Piatti2021} (SMC), photometric centers (maroon “P”) from \citet{vandermarel2001a} (LMC) and \citet{Mateo1998} (SMC), H \textsc{i} kinematic centers (magenta “H”) from \citet{Kim1998} (LMC) and \citet{Stanimirovic2004} (SMC), and—for the LMC—the \citet{Rathore2025} center (orange “R”) and the \cite{Vijayasree2025} center (purple "V"). Table~\ref{tab:MC_centers} gives a summary of these various centers.} 
% \label{fig:MC_centers}
% \end{center}
% \end{figure*}

\begin{figure*}
\begin{center}
\includegraphics[width=1.0\textwidth]{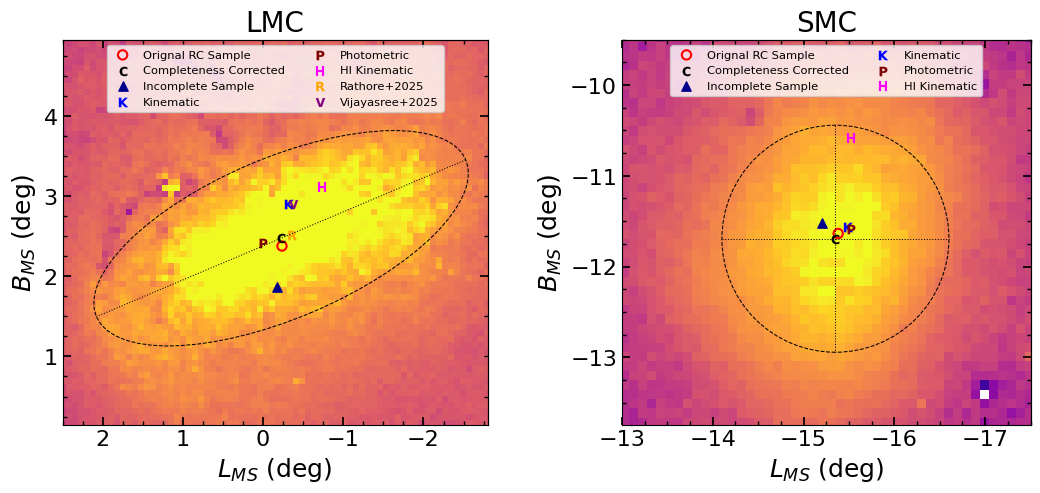}
\caption{\textit{\textbf{Left:}} Center-of-mass estimates for the LMC. \textit{\textbf{Right:}} Same as left, but for the SMC. The centers derived from our original RC samples are shown as red open circles. Completeness-corrected samples are indicated by black “C” markers, and incomplete samples by dark blue triangles. Literature estimates are overplotted for comparison: stellar kinematic centers (blue “K”) from \citet{Choi2022} (LMC) and \citet{Piatti2021} (SMC); photometric centers (maroon “P”) from \citet{vandermarel2001a} (LMC) and \citet{Mateo1998} (SMC); H\,\textsc{i} kinematic centers (magenta “H”) from \citet{Kim1998} (LMC) and \citet{Stanimirovic2004} (SMC); and—for the LMC—the \citet{Rathore2025} center (orange “R”) and the \citet{Vijayasree2025} center (purple “V”). Table~\ref{tab:MC_centers} summarizes all center estimates.}
\label{fig:MC_centers}
\end{center}
\end{figure*}

% Section #3.4
\subsection{RC Dynamical Center}
\label{sec:photometric_center}

Anchoring the center of mass of each Cloud is essential for many aspects of our analysis. The adopted center directly affects all geometric measurements and also provides insight into potential offsets between different tracers—such as differences between the stellar photometric center, the stellar kinematic center, and the H\textsc{i} gas dynamical center \citep[e.g.,][]{Kim1998,Mateo1998,vandermarel2001a,Stanimirovic2004,Piatti2021,Choi2022}. In particular, Sections~\ref{sec:inc_Pos} and~\ref{sec:LMC_warp} require a well-defined center when determining the LMC’s disk inclination, position angle, and azimuthal warp structure. Given the size and homogeneity of our RC sample, we therefore derive updated stellar dynamical centers for both the LMC and SMC.

We employ the iterative shrinking–sphere method of \citet{Power2003} to determine the stellar center of mass of each Cloud. In this approach, we begin by enclosing all RC stars within an initial sphere of radius $10^{\circ}$ on the sky. We compute the mean center of mass (mean $\alpha$ and mean $\delta$) of the stars inside this sphere and then shrink the radius by 30\% at each iteration. At each step, the center of mass is recomputed using only the stars that fall within the shrunken sphere. This procedure suppresses contamination from asymmetric outer structure, tidal debris, and low–density periphery stars, causing the estimate to converge toward the densest and most symmetric region of the stellar distribution. Iterations continue until the shift in the center between successive steps falls below the equivalent of 0.01 kpc at the distance of each Cloud.

To assess the impact of crowding and completeness on the inferred centers, we apply this method to three RC samples for both the LMC and SMC: (1) the original RC samples 
defined in Section~\ref{sec:RC_selection}, (2) the completeness–corrected samples weighted by the inverse of the completeness from Section~\ref{sec:completeness}, and (3) the incomplete samples defined by the $C^{*}$ selection. This comparison allows us to quantify the sensitivity of the derived dynamical centers to completeness effects.

Figure~\ref{fig:MC_centers} presents the results of this analysis alongside a range of literature center of mass determinations for both the LMC and SMC, with a summary provided in Table~\ref{tab:MC_centers}. For both galaxies, the centers derived from our original and completeness–corrected RC samples agree to within $\sim0.1^{\circ}$ for the LMC and $\sim0.07^{\circ}$ for the SMC. In contrast, the incomplete samples introduce significantly larger shifts in the inferred centers, especially in the LMC, consistent with the findings of \citet{Rathore2025}. Our stellar dynamical centers also show good agreement with independent stellar kinematic and photometric centers, while differing most strongly from the H\textsc{i} kinematic centers, which are known to be influenced by the disturbed gaseous morphology of both Clouds.

Taken together, these results demonstrate that our original RC samples are not significantly biased by crowding–related incompleteness. The excellent agreement between the original and completeness–corrected centers indicates that any residual incompleteness does not measurably distort the underlying spatial distribution of RC stars, and therefore does not affect our subsequent geometric modeling. For the remainder of this analysis, we therefore adopt the original RC samples for both the LMC and SMC.

% \begin{figure*}
% \begin{center}
% \includegraphics[width=1.0\textwidth]{Figures/3Panel_RC_Theory_inner_outer_WEST_CMDs.png}
% \caption{\textit{Left:} Theoretical RC isochrones from \citet{Girardi2001} for four metallicities, with higher metallicity corresponding to redder colors. Curves are color-coded by age in Gyr. At fixed age, more metal-poor RC stars are bluer and intrinsically brighter (i.e., have smaller absolute magnitudes $M_{G}$). At fixed metallicity, RC color remains nearly constant with age except at the youngest ($\lesssim$2 Gyr) and oldest ($\gtrsim$9–10 Gyr) extremes, where age becomes the dominant driver of color rather than metallicity.
% \textit{Middle:} Observed RC locus in the western main disk of the LMC ($R \approx 4.8^{\circ}$).
% \textit{Right:}  Observed RC locus in the western diffuse, low-surface-brightness periphery of the LMC ($R \approx 7.4^{\circ}$). In both panels, the ${[\mathrm{Fe/H}]} = -0.69$ dex theoretical RC curve is over plotted and color-coded by age.} 
% \label{fig:RC_theory_curves}
% \end{center}
% \end{figure*}

\begin{figure}
% \begin{center}
\includegraphics[width=0.5\textwidth]{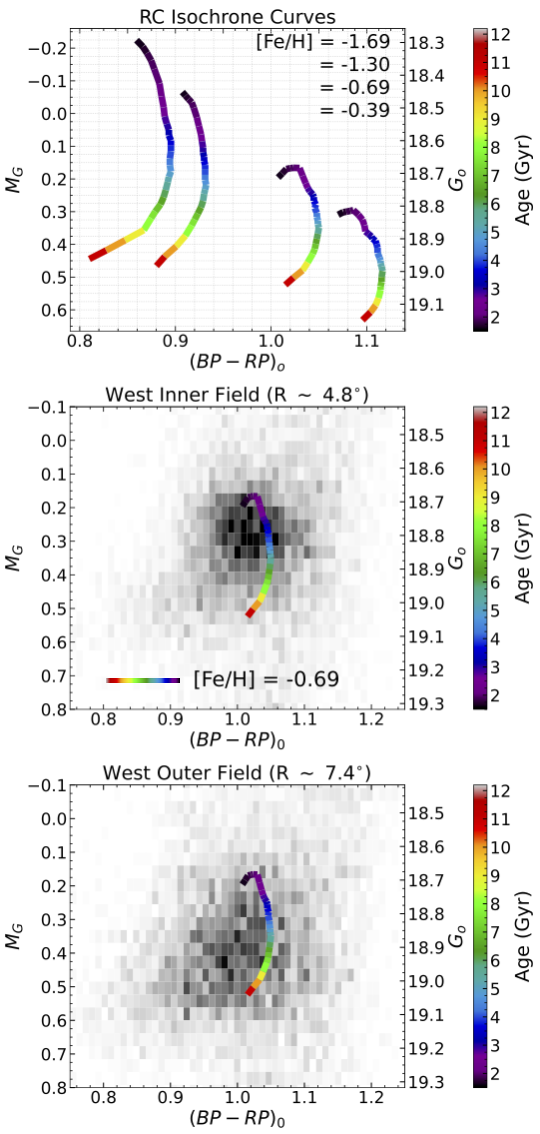}
\caption{\textit{\textbf{Top:}} Theoretical RC isochrones from \citet{Girardi2001} for four metallicities, with higher metallicity corresponding to redder colors. Curves are color-coded by age in Gyr. At fixed age, more metal-poor RC stars are bluer and intrinsically brighter (i.e., have smaller absolute magnitudes $M_{G}$). At fixed metallicity, RC color remains nearly constant with age except at the youngest ($\lesssim$2 Gyr) and oldest ($\gtrsim$9–10 Gyr) extremes, where age becomes the dominant driver of color rather than metallicity.
\textit{\textbf{Middle:}} Observed RC locus in the western main disk of the LMC ($R \approx 4.8^{\circ}$).
\textit{\textbf{Bottom:}} Observed RC locus in the western diffuse, low-surface-brightness periphery of the LMC ($R \approx 7.4^{\circ}$). In both panels, the ${[\mathrm{Fe/H}]} = -0.69$ dex theoretical RC curve is over plotted and color-coded by age.} 
\label{fig:RC_theory_curves}
% \end{center}
\end{figure}

% Section #3.5
\subsection{RC Population-Driven Systematics \& Their Mitigation}
\label{sec:population_systematics}

Two population-driven systematics must be accounted for when deriving distances to the MCs from RC stars: \textit{(i)} spatial variations in metallicity and \textit{(ii)} variations in age (and, by extension, stellar mass). Both vary across the Clouds and change the intrinsic RC luminosity, directly biasing distance estimates if uncorrected.

Variations in metallicity and age imprint differently on the dereddened
$(BP-RP)_0$ color and on the \textit{Gaia} $G$-band absolute magnitude $M_G$.
The left panel of Figure~\ref{fig:RC_theory_curves} illustrates this with four RC isochrones from \citet{Girardi2001} that span a range of metallicities and are color–coded by age. Metal-poor RC stars are intrinsically
brighter and bluer than their metal-rich counterparts,
whereas increasing age primarily makes RC stars progressively fainter while
leaving the color nearly constant—except at the very youngest
($\lesssim 2$~Gyr) and oldest ($\gtrsim 9$~Gyr) epochs, where age also induces a measurable color shift.

% Paragraph #8
% This theoretical expectation is examined observationally in the middle and right panels of Figure~\ref{fig:RC_theory_curves}. Shown is the RC locus for inner (R $\sim$ 4.8$^{\circ}$) and outer (R $\sim$ 7.4$^{\circ}$) LMC fields with the [Fe/H]=$-$0.69 dex RC isochrone curve over plotted. The inner field lies just inside the boundary of the main LMC disk, while the outer field samples the low-surface-brightness diffuse periphery just outside the main LMC disk. The theoretical curves are not intended to represent precise models for the RC populations in each field, but instead serve as qualitative guides for comparison and expectations.The inner field exhibits a prominent young RC population ($\sim$2--4~Gyr), seen as the overdensity at $M_{G}$ $\sim$0.3~mag or $G_{0}$ $\sim$18.8~mag, whereas the outer field is dominated by older RC stars ($\gtrsim 5$~Gyr). These contrasting CMDs illustrate systematic differences in RC age and metallicity between the inner LMC disk and its diffuse outer regions.

% Paragraph #8 (rewritten)
This theoretical expectation is tested observationally in the middle and bottom panels of Figure~\ref{fig:RC_theory_curves}. For reference, we display both apparent magnitudes and absolute magnitudes derived assuming a constant LMC distance modulus of 18.5 mag \citep{deGrijs2014,Pietrzynski2019}. Shown is the RC locus for an inner field (R $\sim$ 4.8$^{\circ}$) and outer field (R $\sim$ 7.4$^{\circ}$), together with the [Fe/H] = $-$0.69 dex RC isochrone over plotted. The inner field lies just inside the boundary of the main LMC disk, while the outer field probes the diffuse, low–surface-brightness periphery just beyond the LMC disk. These theoretical curves are not intended as precise models of the RC population in each region, but rather as qualitative guides for comparison. In the inner field, stars cluster prominently near $M_{G} \sim 0.25$ mag (or $G_{0} \sim 18.8$ mag), consistent with a dominant young RC population of $\sim$2–4 Gyr. By contrast, the outer field lacks this strong clustering and instead shows a higher fraction of stars at fainter magnitudes and slightly bluer colors, indicative of an older RC population ($\gtrsim 5$ Gyr). This picture is consistent with the broader literature that the LMC disk predominantly hosts younger populations, whereas the outer periphery is dominated by older stars \citep[e.g.,][]{Gallart2008,Mackey2016,Nidever2019}.

\begin{figure*}
\begin{center}
\includegraphics[width=1.0\textwidth]{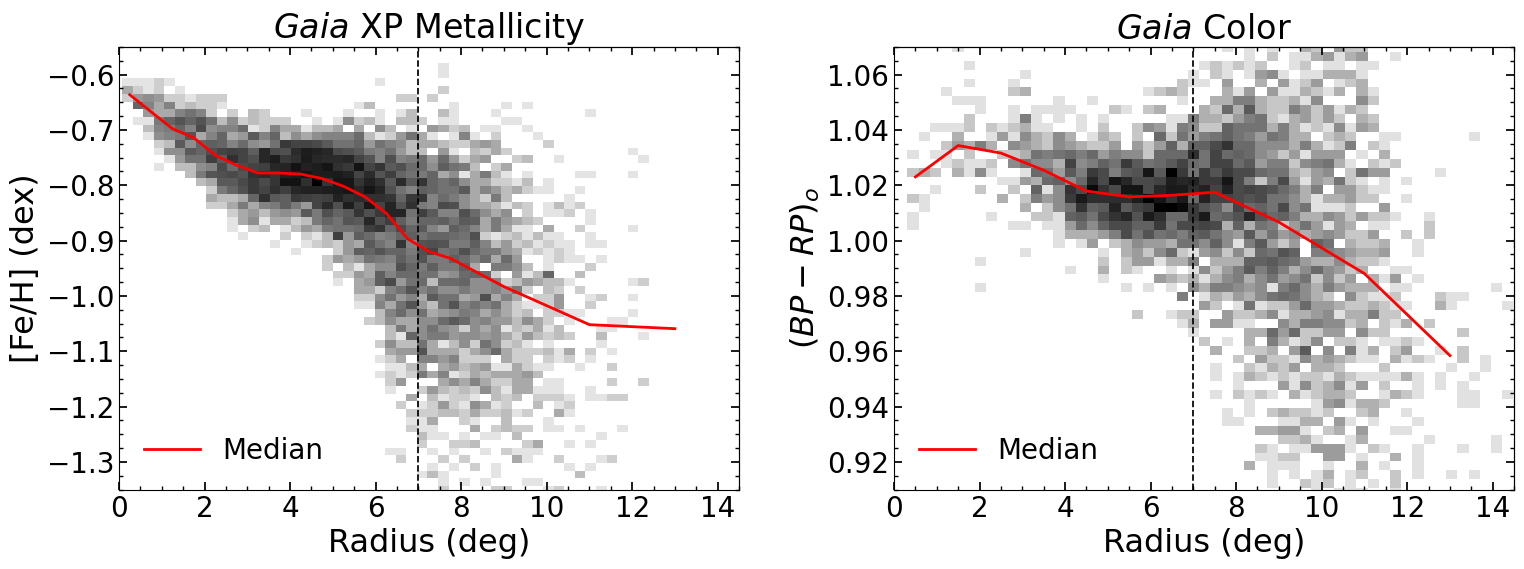}
\caption{\textit{\textbf{Left:}} Global metallicity trend in the LMC using \textit{Gaia} XP metallicities \citep{Andrae2023b},
binned in $10\arcmin \times 10\arcmin$ spatial bins. The red curve shows the median metallicity in each radial bin.
\textit{\textbf{Right:}} Low extinction (i.e., E(B-V) $<$ 0.06) de-reddened \textit{Gaia} $(BP - RP)_{0}$ color of our LMC RC sample, also binned in $10\arcmin \times 10\arcmin$ spatial bins. The red curve shows the median color in each radial bin. In both panels, the approximate extent of the nominal LMC disk is indicated by the dashed black lines at $7^{\circ}$. Excluding the bar region ($\lesssim$2$^{\circ}$), to first order, color traces metallicity.} 
\label{fig:metal_color_vs_radius}
\end{center}
\end{figure*}

Therefore, using a single ``representative'' $M_G$ for all RC stars would therefore bias local distance estimates, particularly in the metal-poor outer LMC and SMC periphery (e.g., the LMC NA or the SMCNOD). Instead,
$M_G$ must be calibrated against an observable proxy, which is most conveniently
the dereddened $(BP - RP)_{0}$ color.

The next sections outline this calibration procedure. In
Section~\ref{sec:MetallicityDependence} we show that metallicity is the dominant
determinant of RC color, enabling the application of the empirical
color--magnitude relation of \citet{Ruiz-Dern2018} to obtain first--order
metallicity--corrected distances (Section~\ref{sec:abs_mag_calibration}).
We then derive spatially resolved SFHs, using these
preliminary distances as priors, to estimate median RC ages and apply a second--order population correction to our derived distances
(Section~\ref{sec:SFH_RC_ages}).

%Applying this method accounts, to first order, for the metallicity dependence of RC luminosity, effectively eliminating the chemical population effect from our results.

% \begin{figure*}
% \begin{center}
% \includegraphics[width=1.0\textwidth]{Figures/2Panel_Gaia_feh_bprp_vs_radius.png}
% \caption{\textit{Left:} Global metallicity trend in the LMC using \textit{Gaia} XP metallicities \citep{Andrae2023b},
% binned in $5\arcmin \times 5\arcmin$ spatial bins. The red curve shows the median metallicity in each radial bin.
% \textit{Right:} Global de-reddened \textit{Gaia} $(BP - RP)_{0}$ color of our LMC RC sample, also binned in $5\arcmin \times 5\arcmin$ spatial bins. The red curve shows the median color in each radial bin. In both panels, the approximate extent of the nominal LMC disk is indicated by the dashed black lines at $7^{\circ}$. Excluding the bar region ($\lesssim$2$^{\circ}$), to first order, color traces metallicity.} 
% \label{fig:metal_color_vs_radius}
% \end{center}
% \end{figure*}

\subsubsection{Metallicity Dependence}
\label{sec:MetallicityDependence}

The left panel of Figure~\ref{fig:metal_color_vs_radius} displays the global trend in \textit{Gaia}~XP metallicities, \([\mathrm{Fe/H}]\), for the LMC \citep{Andrae2023b}\footnote{Updated metallicities from BOSS spectroscopy in the \textit{Magellanic Genesis Survey} suggest a higher mean value of \([\mathrm{Fe/H}] \sim -0.5\ \mathrm{dex}\) 
\citep{Nidever2025b}}. These values are derived from the same dataset used by \citet{Massana2024} but with a slightly different RGB selection. The right panel shows the global, dereddened RC color trend, \((BP - RP)_0\), for our LMC RC sample with $E(B-V) < 0.06$ as a function of radius. In both panels, the red curve marks the median in each radial bin.

The LMC disk exhibits an almost constant metallicity of \(\approx -0.75\ \mathrm{dex}\) out to \(\sim6^{\circ}\).  Within the bar (\(\lesssim2^{\circ}\)), the metallicity rises to \(\approx -0.6\ \mathrm{dex}\), reflecting its younger stellar populations. Correspondingly, \((BP - RP)_0\) color remains nearly flat to \(\sim7^{\circ}\). The slightly bluer colors towards the center ($\lesssim$ $2^{\circ}$) are best explained by the dominance of very young RC stars (\(\lesssim2~\mathrm{Gyr}\)), whose moderate metallicities (\(\sim -0.7\ \mathrm{dex}\)) shift the RC to bluer colors (see Figure~\ref{fig:RC_theory_curves}). This behavior reproduces the SMASH results of \citet{Choi2018a} who see a similar color trend in the southern LMC with a clean sample of RC stars. Beyond \(\gtrsim7^{\circ}\) (delineated by dashed black lines), both the color and metallicity dispersions increase, indicating a larger mix of halo and/or migrated‑disk populations spanning a broader range of ages and metallicities.

Outside the bar‑dominated region (\(\gtrsim2^{\circ}\)), the dereddened \((BP - RP)_0\) color therefore tracks metallicity to first order.  We can thus correct the metallicity dependence of RC luminosity simply by accounting for the observed variation in dereddened color. The same correspondence between RGB metallicities and RC color is also evident in the SMC, confirming that this approach applies to both Clouds.

Although RC and RGB stars occupy different evolutionary phases, homogeneous spectroscopic analyses show that they share indistinguishable metallicity distributions \citep{Bovy2014,Ness2016}. In the MCs, RGB‑based metallicity gradients \citep{Carrera2008a,Carrera2008b,Choudhury2016,Massana2024} mirror the radial trends we observe in RC color, validating the use of RGB metallicities to interpret RC colors (see also \citealt{Girardi2016} for the expected metallicity sensitivity of RC properties).

% Section #3
\subsubsection{RC Absolute Magnitude \& Initial Distance Calculation}
\label{sec:abs_mag_calibration}

%The metallicity–age pairing is itself governed by a galaxy’s SFH and AMR \citep{Girardi1998}.

% Paragraph #1
%To use the RC as a reliable standardized candle, its absolute magnitude must be determined with care.Since RC absolute magnitude is sensitive to both metallicity and age a single ``universal” value can introduce significant systematic uncertainty \citep{Girardi2016}. Consequently, RC absolute magnitude must be calibrated against observable stellar properties, rather than taken to be constant. To first order (excluding the bar region $\lesssim$2$^{\circ}$, as discussed in Section~\ref{sec:RC_pop_effects}), RC color traces metallicity (see Figure~\ref{fig:metal_color_vs_radius}) and therefore approximately the age distribution of the population: the LMC’s inner regions are more metal-rich ({[Fe/H]} $\sim$ –0.7 dex), redder (excluding the bar, see Section~\ref{sec:RC_pop_effects}), and host younger RC stars, whereas its outer periphery is more metal-poor ({[Fe/H]} $\sim$ –1.0 dex), bluer, and dominated by older RC stars.

% Paragraph #2
Guided by the considerations of Sections~\ref{sec:population_systematics} and~\ref{sec:MetallicityDependence}, we adopt the empirical absolute-magnitude–color relation from \citet{Ruiz-Dern2018}, derived using \textit{Gaia} DR1. \citet{Ruiz-Dern2018} assembled a clean sample of 2,482 low-extinction MW RC stars with high-precision parallaxes, ensuring accurate distance estimates. They fit the mode of the RC absolute-magnitude distribution in $G$ as a function of $(G-K_s)$ color, leading to the calibration:

\begin{equation} 
\label{eq:Mg}
M_{G} = 0.495 + 1.121 \cdot (G - K_{s} - 2.1)
\end{equation}

% Paragraph #3  (revised)
Although Equation~\ref{eq:Mg} was calibrated with \textit{Gaia}~DR1 parallaxes, it remains applicable to DR3 photometry because \textit{(i)} the $G$‑band instrumental magnitude is stable across data releases, and \textit{(ii)} the DR3 parallax zero‑point offset lies well within our adopted error budget \citep{Riello2021}. One caveat is that the calibration sample consists of local MW RC stars whose SFH and AMR differ from those of the MCs. On average, Cloud RC stars are more metal‑poor and therefore slightly bluer that the local MW RC population. We nonetheless adopt Equation~\ref{eq:Mg} for three reasons:
\begin{enumerate}
    \item \textbf{Evolutionary phase parity} — Both MW and MC RC stars occupy the same core‑helium‑burning stellar evolutionary phase; the physics encoded in the color term is therefore universal.
    \item \textbf{Color overlap} — The calibration is based on a clean sample of bona‑fide RC stars whose negligible reddened $(G-K_s)$ color range overlaps that of our MC sample, ensuring the relation is evaluated within a similar color domain.
    \item \textbf{Modal fitting robustness} — Fitting the mode of the RC absolute‑magnitude distribution mitigates biases from incompleteness and selection effects, making the relation comparatively insensitive to differences in population mix between the MW and MCs.
\end{enumerate}

% Paragraph #4
To apply Equation~\ref{eq:Mg} to our RC sample, we convert each star’s extinction corrected $(BP - RP)_{o}$ color into $(G - K_s)_{o}$ color using the photometric transformation derived from \textit{Gaia} DR3 photometry \citep{GaiaCollaboration2021a}:

\begin{equation}\label{eq:gks_color}
\begin{split}
(G-K_{s})_{o} &= -0.0981 + 2.089\cdot(BP - RP)_{o} \\
&\quad - 0.1579\cdot(BP - RP)_{o}^{2}
\end{split}
\end{equation}

\noindent Applying this relation, we compute the dereddened $(G-K_s)_{0}$ color and then the absolute magnitude $M_G$ for every RC star in our catalog.  
For the combined MC sample we find a median $(G-K_s)_{0}$ $\approx$ 1.83~mag and a corresponding $M_G$ $\approx$ 0.20~mag.  
Separate medians for each galaxy are  
$(G-K_s)_{0}$ $\approx$ 1.85~mag with $M_G$ $\approx$ 0.22~mag for the LMC, and  
$(G-K_s)_{0}$ $\approx$ 1.77~mag with $M_G$ $\approx$ 0.12~mag for the SMC.

Using the calibrated absolute magnitude relation (Equation~\ref{eq:Mg}) and our final derived extinction corrected photometry described in Section~\ref{sec:extinction}, we calculate first-order metallicity-corrected distances using:

\begin{equation}
\label{eq:dist_modulus}
d~[\mathrm{kpc}] = \frac{10^{(G_{0} - M_{G} + 5)/5}}{1000}
\end{equation}

\noindent This equation follows directly from the definition of the distance modulus and provides distances in kiloparsecs for our entire RC sample. Applying this procedure, we obtain initial median distance estimates of $50.7 \pm 0.01$~kpc to the LMC (within $23^{\circ}$) and $60.9 \pm 0.02$~kpc to the SMC (within $12^{\circ}$). The quoted uncertainties represent purely statistical random errors, estimated through Monte Carlo bootstrapping. In the next section, we describe how independently derived SFHs from DELVE-MC data are used to account for second-order RC population effects in our distance determinations.

\begin{figure}
% \begin{center}
\includegraphics[width=0.5\textwidth]{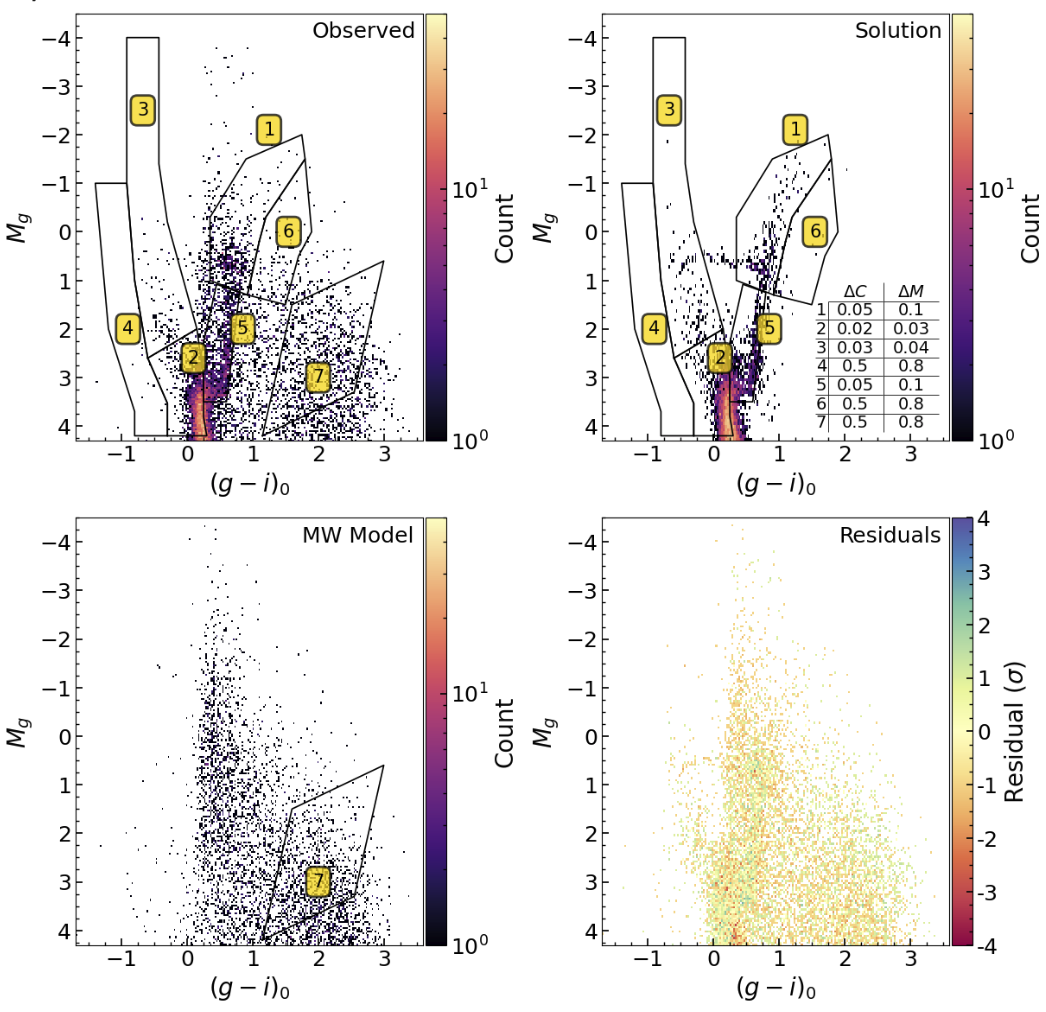}
\caption{\textit{\textbf{Top Left:}} Observed CMD of example DELVE-MC bin in the outer LMC disk, overlaid with our bundle strategy.
\textit{\textbf{Top Right:}} Reproduced best fit \textit{solution} CMD. The inset table shows the binning applied in color ($\Delta$C) and magnitude ($\Delta$M) within each bundle. 
\textit{\textbf{Bottom Left:}} The MW model used for this brick.
\textit{\textbf{Bottom Right:}} The residual CMD (observed - solution) in units of the Poisson uncertainty ($\sigma$).} 
\label{fig:4Panel_SFH}
% \end{center}
\end{figure}

\begin{figure*}
\begin{center}
\includegraphics[width=1.0\textwidth]{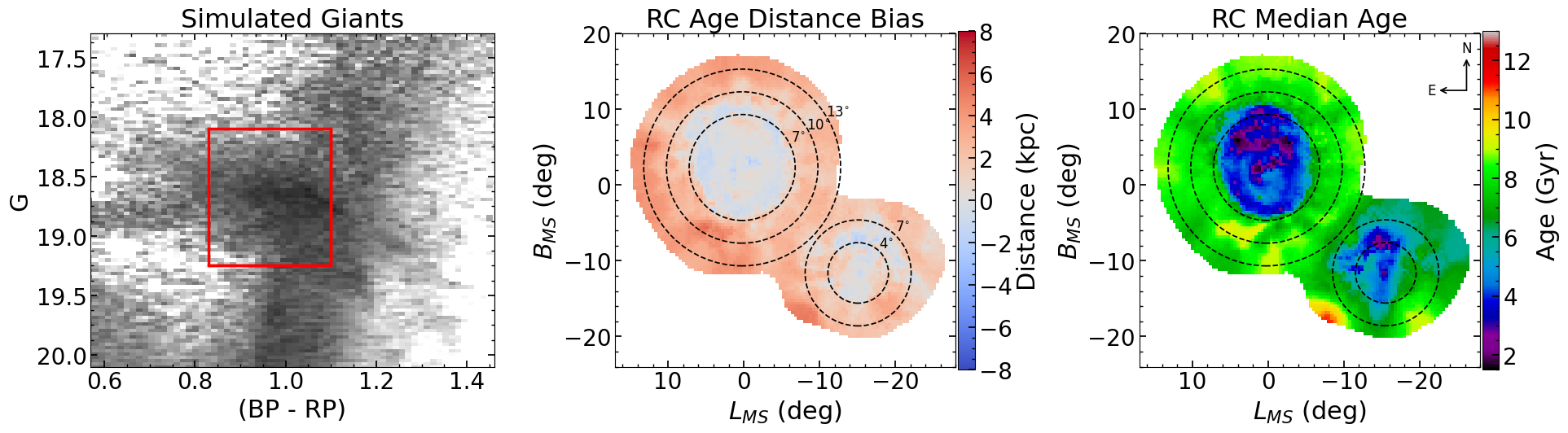}
\caption{\textit{\textbf{Left:}} CMD of synthetic giants generated from the best-fit SFHs across the DELVE-MC footprint. The red box indicates the selection region used to isolate the simulated RC population.  
\textit{\textbf{Middle:}} RC distance bias introduced by variation in RC ages, defined as the difference between the modeled RC distance and the median distance to the sample. Older RC populations appear intrinsically fainter, leading to overestimated distances. The age distance bias is particularly strong in the LMC periphery $\ge$  $7^{\circ}$ where the median RC age is $\sim$ 8.5 Gyrs.
\textit{\textbf{Right:}} Map of the median RC age in Gyr, derived from the SFH solutions in each DELVE-MC bin. RC populations in the outer LMC and SMC are systematically older than those in their inner disks, contributing to the observed distance bias. Dashed circles in the center and right panels mark constant radii from the LMC and SMC centers.} 
\label{fig:3Panel_SimRC_AGEBias_MedAge}
\end{center}
\end{figure*}

% Section #3.5
\subsection{Second-Order Distance Corrections using SFHs}
\label{sec:SFH_RC_ages}

% Paragraph #1
% To account for age-dependent variations in RC luminosity, we use independently derived SFHs (Oden et al., in prep) to estimate the median stellar age throughout the Clouds. These age estimates are then used to derive local distance corrections in our RC sample to our derived distances from Section~\ref{sec:abs_mag_calibration}.

Having corrected first-order population effects by calibrating $M_G$ against color—and thereby accounting for metallicity variations (Sections~\ref{sec:population_systematics}--\ref{sec:abs_mag_calibration})—second-order effects remain due to the age dependence of RC stars, which introduces variations in RC luminosity (see Figure~\ref{fig:RC_theory_curves}). To mitigate this bias, we derive spatially resolved SFHs and use them to estimate the median stellar age across the Clouds, enabling a correction for the age–luminosity dependence of RC stars. The relevant SFH results presented here will be discussed in greater detail, together with a comprehensive analysis of the SFH of the MCs, in \citet{Oden2026b}. In this section, we briefly summarize the SFH methodology and highlight the specific procedures relevant to deriving second-order population-corrected distances.

% To account for age-dependent variations in RC luminosity, we use independently derived SFHs \citealt{Oden2025} to estimate the median stellar age throughout the Clouds. These age estimates are then used to derive local distance corrections to the RC-based distances obtained in Section~\ref{sec:abs_mag_calibration}.  
% As has been shown in prior SFH studies, RC morphology in synthetic CMDs naturally reflects the age and metallicity distributions recovered in the fit \citep[e.g.][]{Monelli2010a,RuizLara2018}, making it possible to quantify RC population effects directly from the SFH solution.

% maybe say something in paragraph 1 how the RC morphology has been previously estimated/constrained in SFH solutions. 

We derive SFHs using deep, multi-band DECam DELVE–MC \citep{Nidever2025a} photometry in the $g$ and $i$ filters. SFH solutions are obtained with the \verb|TheStorm| software \citep{Bernard2015,Bernard2018}, which fits model CMDs to the observed data under the assumption that it is well-represented by a superposition of simple stellar populations (SSPs) of a variety of ages and metallicities. The model CMDs are generated from a compilation of SSPs extracted from a “mother CMD” produced with BaSTI stellar evolutionary models \citep{Pietrinferni2004,Cassisi2006}, spanning ages from 0.02 to 13.9 Gyr and metallicities in the range $-2.27 \le \mathrm{[Fe/H]} \le 0.12$. Following the methodology of previous SFH studies \citep[e.g.,][]{Bernard2015,Bernard2018,RuizLara2020,Sakowska2024}, we employ ``bundles'' to partition each CMD for fitting (see polygons in Figure~\ref{fig:4Panel_SFH}). This approach has been demonstrated to be effective in earlier SFH analyses \citep{Monelli2010a,RuizLara2018,RuizLara2020,Rusakov2021,Massana2022,Sakowska2024}.

% In uncrowded fields, the sigmoid function provides an accurate description of completeness when validated against ASTs \citep[e.g.,][]{Fleming1995,Rejkuba2005,Bonatto2019}, which we have independently verified against SMASH ASTs. 

% Discuss completeness and uncertainties in deriving SFH
Accurate SFH solutions require deep photometry that reaches beyond the oMSTO, making it essential to account for observational completeness and photometric uncertainties. Similar to Section~\ref{sec:completeness} using \textit{Gaia} data, we model completeness as a sigmoid recovery function, whose shape parameter $\alpha$ is determined independently for each field from the slope of the LF fall-off. 
The 50\% completeness magnitude ($m_{50}$) is estimated from the LF peak, with a fixed offset of 0.322 mag applied to locate the 50\% level. This procedure yields a position-dependent completeness limit in both the $g$ and $i$ bands. The completeness function is then applied to the synthetic CMDs by randomly removing stars according to probabilities drawn from the sigmoid model, ensuring that the simulated catalogs reproduce the recovery fraction of the data.

Photometric uncertainties are modeled by fitting a third-order polynomial to $\log_{10}(\sigma_m)$ as a function of magnitude, using only stars with valid, finite errors. The relation is derived from the median error–magnitude trend in 0.75~mag bins. Each synthetic star is then assigned a Gaussian perturbation in magnitude space with a dispersion given by this model, so that the synthetic CMD reproduces the observed scatter across the full magnitude range.

% Paragraph #2
% We derive SFHs with the \verb|TheStorm| code, which fits synthetic CMDs to the observed CMDs under the assumption of simple stellar populations. This approach has proved effective in earlier work \citep{Monelli2010a,RuizLara2018,RuizLara2020,Rusakov2021,Massana2022,Sakowska2024}. The input photometry comes from deep, multi‑band DELVE–MC observations in the $g$ and $i$ filters. A full discussion of the SFH procedure is beyond the scope of this paper and we therefore refer interested readers to Oden et al. (in prep) that will discuss the full CMD-fitting technique across all DELVE-MC bricks. In brief, we generate synthetic CMDs using BaSTI stellar‑evolution models \citep{Pietrinferni2004,Cassisi2006} that span ages from 0.1 to 13 Gyr and metallicities in the range $-2.5 \le \mathrm{[Fe/H]} \le 0.0$. These models are convolved with the photometric uncertainties and completeness functions measured from the data. Following the methodology of previous SFH studies \citep[e.g.][]{Bernard2015,Bernard2018,Massana2022,Sakowska2024}, we employ the “bundle” technique to partition each CMD in the fitting process. 

% Paragraph #3
For our analysis, we restrict the DELVE–MC data to a footprint of $14^{\circ}$ from the LMC and $12^{\circ}$ from the SMC, and is partitioned into hybrid bins defined by requiring $\geq2500$ lower main-sequence stars in Bundle 2. Where this condition is met, we adopt the native DELVE–MC ``brick'' size ($0.25^{\circ} \times 0.25^{\circ}$), which is sufficient for the dense LMC disk and the inner SMC. Otherwise, Voronoi binning \citep{Cappellari2003} is applied with a target of $\sim2500$ stars in Bundle 2. SFH fitting is then performed independently for all 3662 hybrid bins, adopting the initial median RC-based distance estimate in each bin (Section~\ref{sec:abs_mag_calibration}) to convert apparent magnitudes into absolute magnitudes. The mother CMD and observed CMDs are then compared directly in absolute magnitude space.

To mitigate contamination from foreground MW stars and background galaxies, we apply photometric cuts of DAOPHOT \citep{Stetson1987,Stetson1990,Stetson1994} quality metrics  ($ -1.0 < {\tt sharp} < 1.0$, ${\tt chi} < 0.3$) and Source Extractor \citep{Bertin1996} stellar probability (${\tt prob} > 0.7$), as well as incorporate a representative MW model constructed from five queried DELVE–MC fields in the MC periphery. By scaling each MW field by its counts in the bundles, \verb|TheStorm| selects the optimal combination of the five fields that best reproduces the control MW Bundle 7. Figure~\ref{fig:4Panel_SFH} shows an example periphery field ($R\sim 7^{\circ}$) observed CMD, the corresponding solution CMD, the MW model, and the resulting residuals.

% Need to describe how this is different than SMASH methodology SFH fitting

% Paragraph #4
From the best-fit SFH solution in each hybrid bin, we generate a sample of simulated giants by selecting solution CMD stars with $g < 21$ mag and $0 < (g-i) < 4.0$. Figure~\ref{fig:3Panel_SimRC_AGEBias_MedAge} shows a zoomed in CMD of the selected simulated giants. Because the model does not include reddening (we fit de-reddened photometry) or LOS depth effects, we adopt a simple RC box selection (red box in Figure~\ref{fig:3Panel_SimRC_AGEBias_MedAge}) to isolate the simulated RC population. The simulated RC sample is then analyzed identically to the observed data, yielding RC distances within the model. By subtracting the median model distance, we quantify the RC age distance bias—the shift in distance relative to the median, driven by variations in RC age. The resulting RC age distance bias is shown in the middle panel of Figure~\ref{fig:3Panel_SimRC_AGEBias_MedAge}, while the right panel displays the corresponding median RC ages (in Gyr) derived from the SFH. The LMC is centered near ($L_{\rm MS}$, $B_{\rm MS}$) $\sim$ ($0^{\circ}$, $2^{\circ}$), with concentric circles at $7^{\circ}$, $10^{\circ}$, and $13^{\circ}$. The SMC is centered near ($L_{\rm MS}$, $B_{\rm MS}$) $\sim$ ($-15^{\circ}$, $-12^{\circ}$), with concentric circles at $4^{\circ}$ and $7^{\circ}$.

A comparison between the middle and right panels of Figure~\ref{fig:3Panel_SimRC_AGEBias_MedAge} shows that older RC populations—most prevalent in the outer peripheries—produce larger age distance biases. This effect is especially pronounced in the LMC beyond $\sim$7$^{\circ}$, where the bias reaches an average of $\sim$2.5 kpc. To correct for this, we adjust the distances derived in Section~\ref{sec:abs_mag_calibration} by subtracting the RC age distance bias in each localized nearest-neighbor cell. For regions outside the DELVE-MC radial footprint with SFHs, we adopt the median distance bias measured in the periphery ($>$8$^{\circ}$ from the LMC or $>$7$^{\circ}$ from the SMC). This yields RC distances corrected both for metallicity—using the \citet{Ruiz-Dern2018} color relation—and for age-dependent luminosity variations, enabling accurate distance measurements across the MCs, including their low–surface-brightness diffuse peripheries.

\begin{figure*}
\begin{center}
\includegraphics[width=1.0\textwidth]{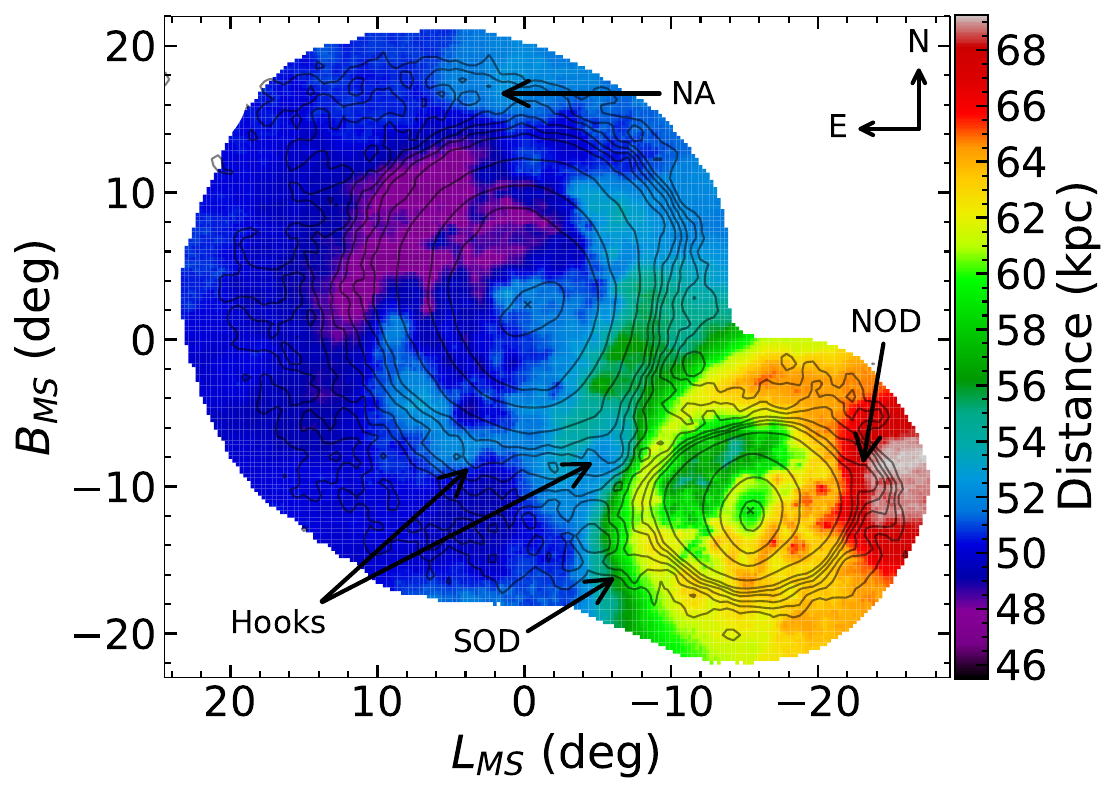}
\caption{Median distance map of the MCs derived from RC stars, corrected for both metallicity and age effects, plotted in MS coordinates. The LMC appears as a large, disk-like structure with a median distance of $\sim$50.6~kpc, while the SMC has a median distance of $\sim$60.7~kpc. The black contour lines trace the RC stellar number density to aid in the visualization of periphery features. Common periphery features are labeled. Black X's indicate the stellar dynamical center for the LMC and SMC. A clear offset between the Clouds is visible, along with hints of structure such as the LMC's outer warp and the SMC's elongation. This distance map is available for public use online \href{https://github.com/slateroden/XMC_DistanceMap}{here}
} 
\label{fig:MC_Distance_Map}
\end{center}
\end{figure*}

% Section #4
\section{Results}
\label{sec:results}

% Section #4.1
\subsection{Distance Map}
\label{sec:MC_maps}

% Paragraph #1 (method)
We present the distance map of the MCs as a spatially smoothed (i.e., averaged) representation rather than a noisy, binned version \footnote{This distance map is available online at \url{https://github.com/slateroden/XMC_DistanceMap}}. We adopt the same nearest-neighbor strategy described in Section~\ref{sec:RC_selection} on a $0.3^{\circ}$ grid, searching for 10,000 stars within a radius of $0.3^{\circ}$. If fewer than 700 stars are found, we increase the search radius to $5^{\circ}$ and take the median of 700 stars. This approach enables us to sample both the high-density regions and the low–surface-brightness periphery.

% Paragraph #2 (map description)
Figure~\ref{fig:MC_Distance_Map} presents the spatially smoothed median RC distance map of the MCs in MS coordinates, while Table~\ref{tab:periphery_distances} summarizes the inferred distances for key periphery features. This map incorporates corrections to systematic RC population effects as described in Sections~\ref{sec:population_systematics}–\ref{sec:SFH_RC_ages}. The map traces the three-dimensional structure of the Magellanic stellar system out to galactocentric radii of ${\sim23^{\circ}}$ ($\sim$20~kpc) from the LMC and ${\sim12^{\circ}}$ ($\sim$10.5~kpc) from the SMC. Black contours trace the RC stellar number density, helping to visualize features such as the LMC Northern Arm (NA), the southern hooks, and the SMCNOD and SMCSOD. For orientation, north–east arrows are included in the upper corner, and major periphery features are labeled directly on the map. The relative distance offset between the Clouds is clearly visible, with a sharp transition near their interface. The inclined geometry of the LMC disk is also apparent, with closer distances toward the northeast and larger distances toward the southwest.

\begin{table*}[ht]
\centering
\caption{Distances of Magellanic Periphery Features}
\label{tab:periphery_distances}
\begin{tabular*}{\textwidth}{@{\extracolsep{\fill}} l r c c c }
\hline\hline
Feature & Number of Stars & $(L_{\mathrm{MS}}, B_{\mathrm{MS}})$ (deg) &  Aperture (deg) & Distance (kpc)  \\
\hline
Northern Arm (base)  & 208  & $(2.8, 17)$    & $R = 1.5$  & $52.8 \pm 0.5$  \\
Northern Arm (end)  & 101  & $(16.5, 16)$    & $R = 1.25$  & $51.1 \pm 1.2$  \\
Southeast Hook    & 305  & $(3.1, -7.5)$   & $R = 1.0$   & $52.1 \pm 0.6$  \\
Southwest Hook    & 284  & $(-6.0,-7.5)$   & $R = 1.0$  & $51.6 \pm 0.4$  \\
Magellanic Bridge & 179  & $(-9.2,-6.3)$   & $R = 1.25$   & $62.4 \pm 1.2$ \\
Eastern SMC     & 623  & $(-10.9,-11.4)$  & $R = 0.5$    & $58.8 \pm 0.7$   \\
SMCNOD       & 735  & $(-23.0,-9.5)$  & $R = 1.5$    & $67.9 \pm 0.7$  \\
SMCSOD       & 328  & $(-7.5,-14.5)$  & $R = 1.5$    & $61.8 \pm 0.8$  \\
\hline
\end{tabular*}
\end{table*}

The LMC appears as a large, elliptical structure centered near $(L_{\mathrm{MS}}, B_{\mathrm{MS}}) \sim (0^\circ, 2^\circ)$, with a median distance of $50.62 \pm 0.01$~kpc measured within $23^{\circ}$. The SMC, centered at $(L_{\mathrm{MS}}, B_{\mathrm{MS}}) \sim (-15^\circ, -12^\circ)$, lies farther away at a median distance of $60.75 \pm 0.02$~kpc within $12^{\circ}$. The quoted uncertainties are purely statistical, derived from Monte Carlo bootstrapping; a detailed assessment of systematic errors is presented in Section~\ref{sec:systematic_errors}. Our results are in excellent agreement with recent high-precision eclipsing binary measurements: the LMC distance is consistent with the $49.59$ $\pm$ $0.54$~kpc estimate of \citet{Pietrzynski2019}, while the SMC distance is only slightly smaller than the $62.4 \pm 0.81$~kpc reported by \citet{Graczyk2020}, with both values within our systematic error budget. Importantly, our wide areal coverage spans the full intermediate-age extent of both Clouds, including their diffuse peripheries, yielding global median distances that are not ``biased'' toward the dense inner regions.

% Paragraph #4 (LMC Periphery)
% The periphery of the LMC is highly substructured. The LMC NA, with its base near $(L_{\mathrm{MS}}, B_{\mathrm{MS}}) \sim (3^\circ, 17^\circ)$, has a median distance of $52.8 \pm 0.5$~kpc at its base and $51.6 \pm 1.2$~kpc toward the end of the feature\footnote{We define the “end” of the feature near $(L_{\mathrm{MS}}, B_{\mathrm{MS}}) \sim (16.5^\circ, 16^\circ)$ but do not claim this is its true termination.}, corresponding to a total difference of $1.2 \pm 1.2$~kpc or a gradient of $-0.1 \pm 0.09$~kpc~deg$^{-1}$. This NA distance gradient is smaller than that reported by \citet{Cullinane2022a}, but we nonetheless detect a small negative distance gradient along the feature. Similarly our median distance to the NA base is slightly closer than the $\sim$57~kpc reported by \citet{elyoussoufi2021}, a difference we attribute to our correction for peripheral age effects. Interestingly, the NA base lies $\sim$3.7~kpc farther away than the surrounding periphery just interior to the feature, suggesting that its true origin may be obscured by projection effects.

% Paragraph #4 (LMC Periphery)
The periphery of the LMC is highly substructured. The LMC NA, anchored near 
\((L_{\mathrm{MS}}, B_{\mathrm{MS}}) \sim (3^\circ, 17^\circ)\), has a median distance of 
\(52.8 \pm 0.5\)~kpc at its base and \(51.6 \pm 1.2\)~kpc toward its tip,\footnote{We define 
the ``end'' of the feature near \((L_{\mathrm{MS}}, B_{\mathrm{MS}}) \sim (16.5^\circ, 16^\circ)\), 
though we do not claim this marks its true termination.} corresponding to a total difference 
of \(1.2 \pm 1.2\)~kpc, or a gradient of \(-0.1 \pm 0.09\)~kpc~deg\(^{-1}\). This measured 
gradient is smaller than that reported by \citet{Cullinane2022a}, but still indicates a weak 
negative distance trend along the structure. Our median distance to the NA base is also closer than the \(\sim 57\)~kpc value found by \citet{elyoussoufi2021}, a discrepancy 
we attribute to our explicit correction for age-dependent luminosity variations. 
Notably, the NA base lies \(\sim 3.7\)~kpc farther away than the adjacent periphery immediately 
interior to the feature, hinting that its apparent morphology may be influenced by projection 
effects, a feature that should be examined further in simulations.

% Paragraph #5 (LMC Periphery Continued)
The two southern hooks—the southeastern (SE) and southwestern (SW) features—are located near $(L_{\mathrm{MS}}, B_{\mathrm{MS}}) \sim (3.0^\circ, -7.5^\circ)$ and $(-6.0^\circ, -7.5^\circ)$, with median distances of $52.1 \pm 0.6$~kpc and $51.6 \pm 0.4$~kpc, respectively. These values are slightly closer than the $\sim$53–55~kpc reported by \citet{elyoussoufi2021}. We argue that our values more accurately reflect the true distances due to the inclusion of metallicity and age corrections. In addition, more distant azimuthal substructures are visible just beyond the main LMC disk, indicative of a fully developed peripheral warp (see Section~\ref{sec:LMC_warp} for a full discussion of the LMC peripheral warp).

% Paragraph 6 (SMC periphery)
% The SMC’s outskirts, though smaller in extent, also shows prominent substructures. The SMCNOD, near $(L_{\mathrm{MS}}, B_{\mathrm{MS}}) \sim (-23.0^\circ, -9.5^\circ)$, has a median distance of $67.9 \pm 0.7$~kpc, slightly further away than the $65.4 \pm 1.1$~kpc reported by \citet{elyoussoufi2021}. A diffuse stellar feature northeast of the SMCNOD, referred to as the “SMC leading feature” by \citet{Massana2024}, lies near $(L_{\mathrm{MS}}, B_{\mathrm{MS}}) \sim (-22.0^\circ, -5.0^\circ)$ at a median distance of $66.5 \pm 1.0$~kpc. Its distance and metallicity distribution (see Figure 6 of \citealt{Massana2024}) closely resembles that of the SMC, suggesting that it consists of stripped or perturbed SMC stellar material. 

% Paragraph 6 (SMC periphery)
The outskirts of the SMC, though less extended than  the LMC, also exhibit prominent substructures. The SMCNOD, located near \((L_{\mathrm{MS}}, B_{\mathrm{MS}}) \sim (-23.0^\circ, -9.5^\circ)\), has a median distance of \(67.9 \pm 0.7\)~kpc—slightly farther than the \(65.4 \pm 1.1\)~kpc reported by \citet{elyoussoufi2021}. A diffuse stellar overdensity northeast of the SMCNOD, identified as the ``SMC leading feature'' by \citet{Massana2024}, lies near \((L_{\mathrm{MS}}, B_{\mathrm{MS}}) \sim (-22.0^\circ, -5.0^\circ)\) with a median distance of \(66.5 \pm 1.0\)~kpc. Its distance and metallicity distribution (see Figure~6 of \citealt{Massana2024}) closely match those of the SMC itself, supporting the interpretation that it represents tidally stripped or dynamically perturbed SMC stellar material.

% Paragraph 7 (SMC periphery continued)
The SMCSOD, discussed by \citet{Massana2024} and located near $(L_{\mathrm{MS}}, B_{\mathrm{MS}}) \sim (-7.5^\circ, -14.5^\circ)$, has a median distance of $61.8 \pm 0.8$~kpc. To our knowledge, this is the first distance estimates for the feature. We also confirm a distance bimodality in the eastern SMC, with two populations averaging $\sim$53~kpc and $\sim$62~kpc. These values agree with previous studies that have investigated this feature in greater detail \citep[e.g.,][]{Hatzidimitriou1989, Nidever2013, elyoussoufi2021, Omkumar2021}. In addition, we detect a global distance gradient toward larger distances across the SMC in the northern direction (toward more negative $L_{\mathrm{MS}}$ values), consistent with \citet{Omkumar2021} findings.

% Closing paragraph (transition)
In summary, our distance map highlights both the large-scale geometry of the Clouds—including the relative offset between the LMC and SMC and the inclined structure of the LMC disk—as well as a wealth of substructures in their peripheries, such as the NA, southern hooks, SMCNOD, and SMCSOD. These results underscore the complex three-dimensional morphology of the Magellanic system and provide the foundation for quantifying the LMC’s disk geometry. In the following section, we constrain the inclination and position angle of the LMC disk by fitting its projected structure as a function of radius.

\subsection{Inclination ($i$) and Position Angle ($\theta$) of LMC}
\label{sec:inc_Pos}

% Paragraph #1 (intro and background)
In this section, we constrain the geometry of the LMC disk by fitting the inclination angle, $i$, and the position angle of the line of nodes (LON), $\theta$, for our RC sample as a function of galactocentric radius. The inclination angle is defined as the tilt between the LMC disk plane and the plane of the sky, with positive inclination indicating that the disk is tipped toward the observer. The position angle (PA) of the LON corresponds to the angle of intersection between the disk plane and the sky plane, measured east of north by convention. For a detailed discussion of the LMC’s disk geometry, see \citet{vandermarel2001a}, \citet{vandermarel2002}, 
\citet{Choi2018b}, and
\citet{JimenezArranz2023}.

% Gaia EDR3 LMC center
% (ra,dec) = (81◦.01,−69◦.38)

\begin{figure*}
\begin{center}
\includegraphics[width=1.0\textwidth]{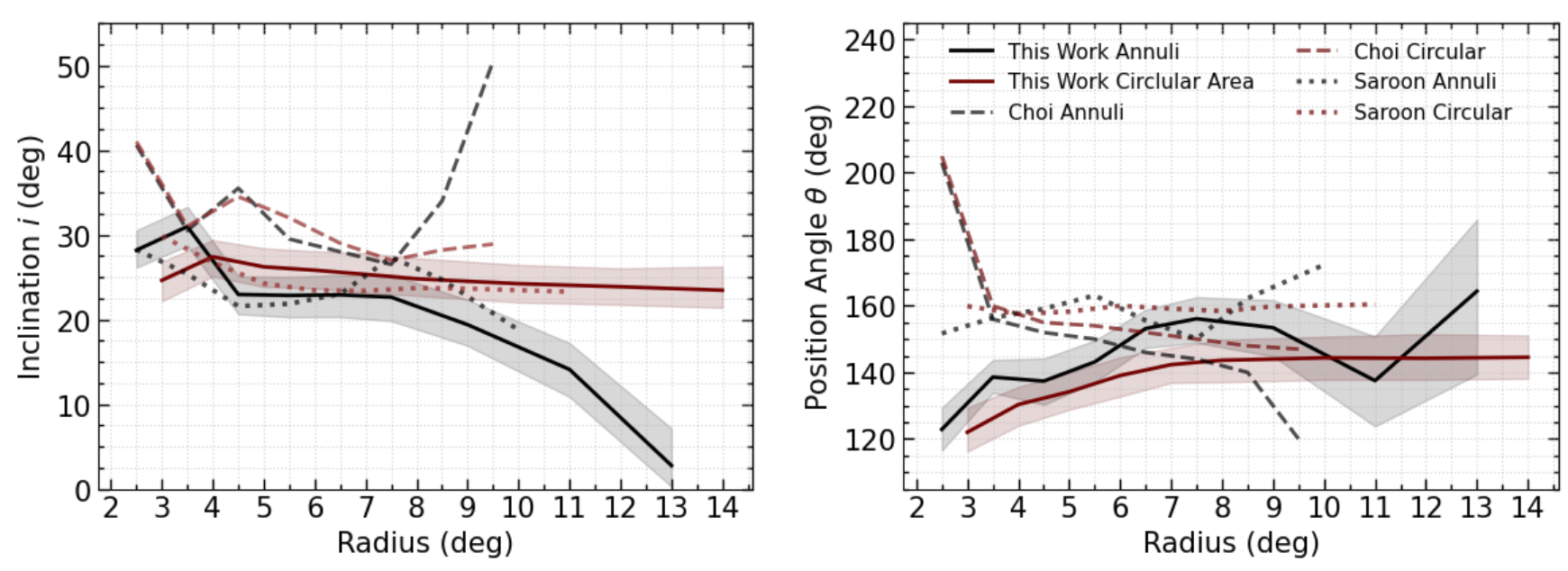}
\caption{Radial variation of the LMC disk geometry.
\textit{\textbf{Left:}} Inclination angle $i$ as a function of galactocentric radius from the LMC center (in degrees). \textit{\textbf{Right:}} Position angle of the line of nodes (LON), $\theta$, as a function of the same radius. Black solid lines show measurements from this work using elliptical annuli: 1$^\circ$ wide from 0$^\circ$–8$^\circ$ and 2$^\circ$ wide from 8$^\circ$–14$^\circ$. Red solid lines indicate circular-area fits at each radius. The black (red) shaded regions denote the 1$\sigma$ uncertainties for our reported $i$ ($\theta$) values. For comparison, previously published measurements from \citet{Choi2018a} are over-plotted as dashed lines, and those from \citet{Saroon2022} as dotted lines, using both annular and circular fitting approaches as labeled.}
\label{fig:inc_theta}
\end{center}
\end{figure*}

% Paragraph #2 (method)
To characterize the disk plane, we convert each star’s equatorial coordinates ($\alpha$, $\delta$) and distance ($D$) into an LMC-centered Cartesian frame $(X, Y, Z)$ following \citet{vandermarel2001a}. We adopt as the LMC center our newly derived stellar dynamical center at ($\alpha_{0}$, $\delta_{0}$) = ($80.60^{\circ}$, $-69.80^{\circ}$) and set the fiducial distance to $D_{0} = 49.15$~kpc, corresponding to the median RC distance within the inner $0.01^{\circ}$ of our sample. Our results are insensitive to modest changes in $D_{0}$ (of order $\pm$1~kpc), and the adopted value is fully consistent with the eclipsing-binary distance from \citet{Pietrzynski2019}.

We also tested two commonly used centers—($\alpha_{0}$, $\delta_{0}$) = ($81.90^{\circ}$, $-69.87^{\circ}$) from \citet{vandermarel2001a} and ($\alpha_{0}$, $\delta_{0}$) = ($80.44^{\circ}$, $-69.27^{\circ}$) from \citet{Choi2022}—and find that our inferred disk geometry remains robust to these choices within our 1$\sigma$ uncertainties. In this coordinate system, the $X$-axis is anti-parallel to right ascension, the $Y$-axis is parallel to declination, and the $Z$-axis points toward the observer.

% Radial variation of the LMC disk geometry.
% The left panel shows the inclination angle $i$, and the right panel shows the position angle (PA) of the line of nodes (LON) $\theta$, both plotted as a function of galactocentric radius from the LMC center (in degrees). Black solid lines represent measurements from this work using elliptical annuli: 1$^\circ$ wide from galactocentric radii 0$^\circ$ to 8$^\circ$, and 2$^\circ$ wide annuli from galactocentric radii 8$^\circ$ to 14$^\circ$. Red solid lines indicate circular area fits at each radius. The black (red) shaded regions show our 1$\sigma$ uncertainty for our reported $i$ ($\theta$) values.
% For comparison, previously published measurements from \citet{Choi2018a} are shown as dashed lines, and from \citet{Saroon2022} as dotted lines, using both annular and circular fitting approaches as labeled.

% Paragraph #3 (method cont.)
We fit a plane to the (X, Y, Z) positions of all stars by minimizing their orthogonal distance to the plane using linear least-squares regression. The plane is defined as:

\begin{equation}
Z = A X + B Y + C
\end{equation}

\noindent From our derived best-fit parameters $(A,\ B,\ C)$, we calculate ($i$, $\theta$) as follows:

\begin{subequations}
\begin{equation}\label{eq:inc}
i = \arccos\left( \frac{1}{\sqrt{A^2 + B^2 + 1}} \right)
\end{equation}

\begin{equation}\label{eq:LON}
\theta = \arctan\left( \frac{-A}{B} \right) + \operatorname{sign}(B) \cdot \frac{\pi}{2}
\end{equation}
\end{subequations}

% Paragraph #4 (global i and theta values)
For the inner LMC disk ($R < 7^{\circ}$), we find a best-fit inclination of $i = 25.32^{\circ} \pm 0.10^{\circ}$ and a LON position angle of $\theta = 142.34^{\circ} \pm 0.21^{\circ}$. However, the disk is known to contain twists and warps \citep{JimenezArranz2025}, which can bias global fits. Restricting the analysis to a more regular portion of the disk ($2^{\circ} < R < 4^{\circ}$; excluding the bar) yields $i = 29.29^{\circ} \pm 0.15^{\circ}$ and $\theta = 137.67^{\circ} \pm 0.33^{\circ}$. Reported uncertainties are purely statistical, derived from a Monte Carlo bootstrapping analysis of the plane-fitting procedure. These values of $(i, \theta)$ are in good agreement with previous determinations \citep[e.g.,][and references therein]{Choi2018a, Saroon2022} .

%Paragraph #5 
Previous studies have shown that the LMC disk is not a flat, uniform structure, but instead exhibits pronounced warping, twisting, and flaring—features likely induced by tidal interactions with both the SMC and the MW \citep[e.g.,][]{vandermarel2001a, Olsen2002, Nikolaev2004, Subramanian2009b, Balbinot2015, JimenezArranz2025}. Consequently, both $i$ and $\theta$ vary with galactocentric radius and depend on spatial coverage, tracer population, and measurement method. One of the earliest wide-field RC-based structural studies, \citet{Olsen2002}, fit a tilted plane to RC magnitudes and derived an inclination of $i = 35.8^{\circ} \pm 2.4^{\circ}$ and a LON position angle of $\theta = 145^{\circ} \pm 4^{\circ}$. However, they also identified a set of outlier fields in the southwest where the RC appeared $\sim$0.1 mag brighter than predicted by the fitted plane, which they interpreted as evidence of a warp, with vertical displacements reaching up to $\sim$2.5 kpc.

%Paragraph #6
Subsequent studies have confirmed and extended these results. \citet{Choi2018a} found that $i$ and $\theta$ vary gradually with radius in the outer disk but more rapidly in the inner regions. Using \textit{Gaia} EDR3 data, \citet{Saroon2022} obtained results consistent with those of \citet{Choi2018a}, reinforcing the presence of radial variations in disk orientation. Together, these studies establish that the LMC disk is significantly warped and twisted, with its geometry evolving as a function of radius.

%Paragraph #7
Figure~\ref{fig:inc_theta} presents our derived values of ($i$, $\theta$), computed using Equations~\ref{eq:inc} and~\ref{eq:LON}, as a function of LMC galactocentric radius. The black solid lines show measurements from annular fits in $1^{\circ}$-wide bins spanning $0^{\circ}$–$8^{\circ}$ and $2^{\circ}$-wide bins spanning $8^{\circ}$–$14^{\circ}$. The red solid lines show results from circular fits within growing apertures at each indicated radius. The shaded regions represent the $1\sigma$ uncertainties, derived using Monte Carlo bootstrapping of the annular (black) and circular (red) fits. By construction, the annular fits provide a more localized view of the disk geometry, while the circular fits capture the global orientation within each aperture. For comparison, the black dashed (dotted) lines show the annulus-based measurements from \citet{Choi2018a} \citep{Saroon2022}, while the red dashed (dotted) lines show their circular-fit results. For consistency with these earlier works, we exclude the central bar region ($R \leq 2^{\circ}$) from our analysis.

% Paragraph #8
Consistent with previous studies, we find a strong radial dependence of ($i$, $\theta$). In particular, the inclination values derived from circular-area fits increase modestly within the inner $4^{\circ}$ and then remain nearly constant at $\sim$26$^{\circ}$. This flat behavior at larger radii agrees well with \citet{Saroon2022}, while the asymptotic $i$ value agrees well with the final circular-area measurement of \citet{Choi2018a}. The rising trend in the inner disk, however, differs from both \citet{Choi2018a} and \citet{Saroon2022}. We note that a similar increase in $i$ within $R < 4^{\circ}$ has been reported by \citet{vandermarel2001a} and more recently by \citet{JimenezArranz2025}, likely associated with the presence of the bar.

% Paragraph #9
Beyond the inner disk ($R > 4^{\circ}$), our inclination values from annular fits show a declining trend, decreasing from $\sim$30$^{\circ}$ to $\sim$4$^{\circ}$ at $R = 13^{\circ}$. Our annular measurements diverge from those of \citet{Choi2018a} beyond $7^{\circ}$, likely due to differences in spatial coverage. In contrast, they are in good agreement with the declining $i$ profile reported by \citet{Saroon2022} at these radii.

% Our $\theta$ values derived from circular area fits show an overall flat behavior at all radii with a value of $\sim$150$^{\circ}$. This flat behavior in $\theta$ is in good agreement with \citet{Choi2018a} and \citet{Saroon2022}. Similarly, $\theta$ values from annular fits exhibit only a weak dependence on radius, excluding the innermost region ($<$3$^{\circ}$) and outside the LMC disk ($>$7$^{\circ}$). 

% Paragraph #10 (NEED to rewrite)
% Our position angle measurements, obtained from both circular-area and annular fits, agree closely with those of \citet{Choi2018a} and \citet{Saroon2022} across the main disk ($R < 7^{\circ}$). The circular-area fits yield an almost flat $\theta$ profile at all radii, clustering around $\sim$150$^{\circ}$. The annular fits are likewise nearly constant, except for deviations in the innermost region ($R < 3^{\circ}$) and beyond the disk edge ($R > 7^{\circ}$). Inside $3^{\circ}$, our results diverge from those of \citet{Choi2018a}$,$ but allowing measurements within $1^{\circ}$ of the LMC center reveals a rising trend in $\theta$, consistent with their findings. Between $\sim6^{\circ}$ and $7^{\circ}$, our profile turns over and begins to decline, again matching both comparison studies. Strikingly, beyond $11^{\circ}$ we detect a renewed increase in $\theta$—a feature not previously reported—which may signal an outer-disk twist or warp warranting further investigation.

% Paragraph #10 
Our position angle measurements, obtained from both circular-area and annular fits, are broadly consistent with those of \citet{Choi2018a} and \citet{Saroon2022}. Both fitting methods show an increase in LON PA from $\sim125^{\circ}$ to $\sim150^{\circ}$ across the inner LMC disk, after which the circular profile remains nearly flat at $\sim145^{\circ}$ out to $\sim14^{\circ}$. The annular fits exhibit somewhat larger deviations, capturing local perturbations in the disk structure. This inner-disk rise in LON PA is in good agreement with \citet{JimenezArranz2025}, who found a similar trend within $R \leq 5^{\circ}$ using independent kinematic methods. Beyond $\sim11^{\circ}$, however, our annular measurements reveal a renewed increase in $\theta$—a feature not previously reported—which may signal an outer-disk twist that warrants further investigation.

% Paragraph #11 (summary)
Overall, our derived ($i$, $\theta$) values are broadly consistent with previous measurements, while extending to larger radii. Our global best fits of $i = 25.32^{\circ} \pm 0.10^{\circ}$ and $\theta = 142.34^{\circ} \pm 0.21^{\circ}$ agree with the global disk values reported by \citet{Choi2018a} and \citet{Saroon2022}. These values also fall well within the broad range of literature estimates, spanning $i \sim 7^{\circ}$–40$^{\circ}$ and $\theta \sim 100^{\circ}$–180$^{\circ}$ \citep[e.g.,][]{vandermarel2001a, Olsen2002, Koerwer2009, Subramanian2010, Subramanian2013, Inno2016, JacyszynDobrzeniecka2017,GaiaCollaboration2021b,Kacharov2024}. In particular, our $\theta$ values align closely with those of \citet{Olsen2002}, and our inclination is comparable to the outer northern disk value of $i = 25.18^{\circ} \pm 0.71^{\circ}$ measured by \citet{Mackey2016} through elliptical modeling of stellar number counts.

\subsection{LMC Azimuthal Warp}
\label{sec:LMC_warp}

\begin{figure}
\includegraphics[width=0.5\textwidth]{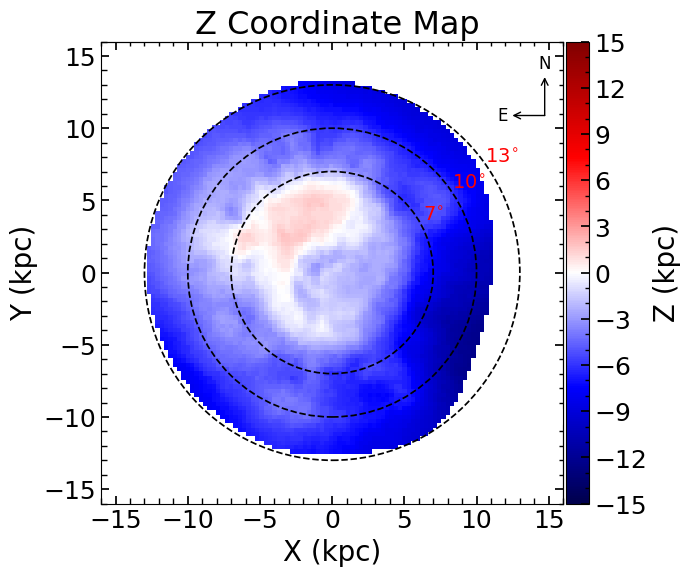}
\caption{Z-coordinate map in the on-sky cartesian frame, color-coded by the vertical coordinate $Z$ (in kpc). The map is centered on our newly derived RC LMC center ($\alpha$,$\delta$) = ($80.60^{\circ}$,$-69.80^{\circ}$) and fiducial distance of $D_{0}$ = 49.15~kpc. Concentric dashed circles indicate angular separations of $7^\circ$ ($\sim$ 6~kpc), $10^\circ$ ($\sim$ 9~kpc) and $13^\circ$ ($\sim$ 11.4~kpc) from the LMC center. A prominent central spiral feature is visible in red, but rapidly dissipates due to the LMC’s disk inclination. At radii beyond $\sim$7$^\circ$, the peripheral warp becomes strikingly apparent in the vertical structure as blue color.} 
\label{fig:Z_Map}
\end{figure}

\begin{figure*}
\begin{center}
\includegraphics[width=1.0\textwidth]{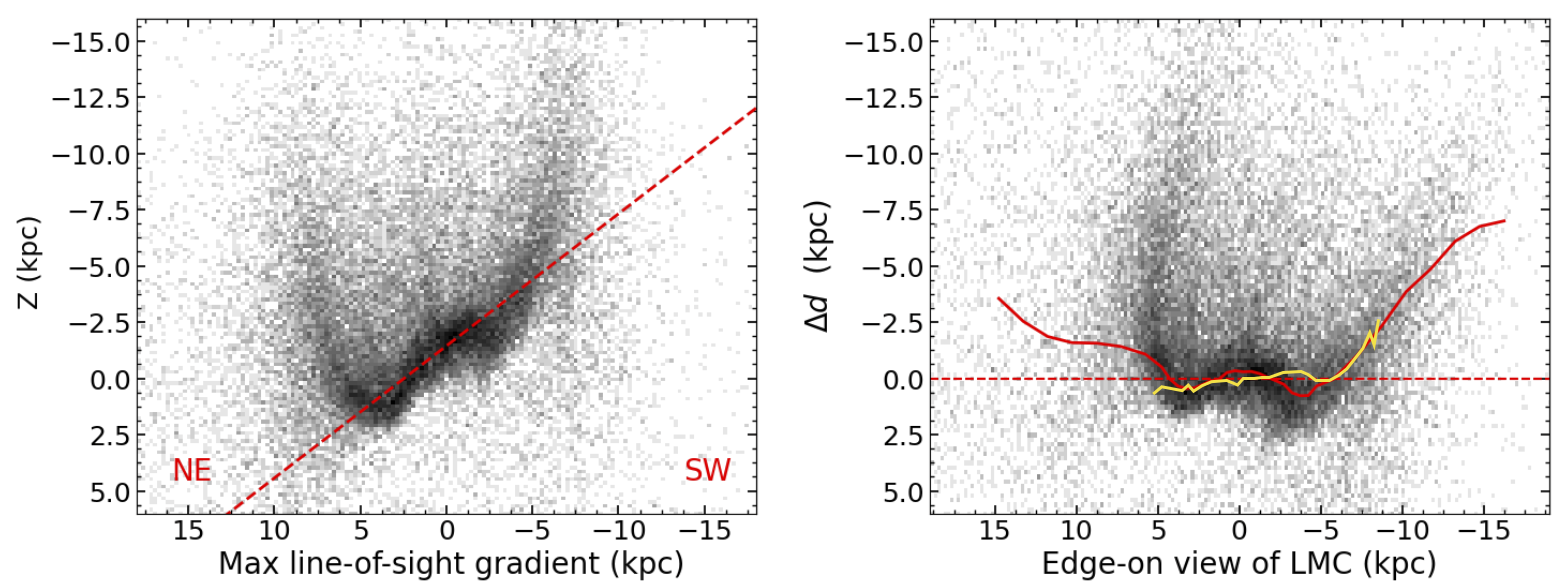}
\caption{Edge-on view of the LMC disk structure.
\textit{\textbf{Left:}} The 3D distribution of LMC RC stars projected along the axis of maximum line-of-sight depth gradient (i.e., perpendicular to the line of nodes), plotted against the vertical coordinate $Z$ (in kpc). The inclination of the LMC disk causes the northeastern (NE) side to be closer and the southwestern (SW) side farther away. The red dashed line shows our inner best-fit plane to the disk, capturing its global tilt.
\textit{\textbf{Right:}} The same edge-on projection after correcting for the disk’s inclination. The vertical axis shows the residual distance $\Delta d$ from the best-fit plane. The red curve traces the median residual distance, weighted by the local on-sky stellar density. The yellow curve shows the median profile from \citet{Choi2018a}, over-plotted for comparison.}
\label{fig:edge_view}
\end{center}
\end{figure*}

\begin{figure*}
\begin{center}
\includegraphics[width=1.0\textwidth]{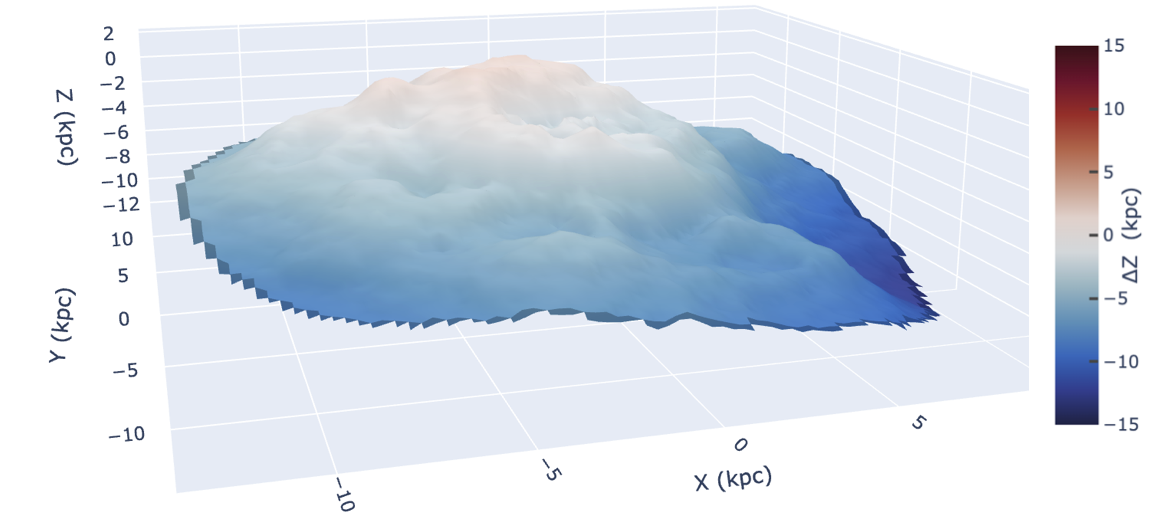}
\caption{3D plot in the on-sky cartesian coordinate system (X,Y,Z) described in Section~\ref{sec:inc_Pos} of our LMC RC sample out to $\sim$13$^{\circ}$. The colored surface is the median Z-coordinate with blue (red) indicating further (closer) distances than the adopted LMC centric distance of $D_{o}$ = 49.15~kpc. An animation of this plot from different viewing perspectives is available online \href{https://slateroden.github.io/lmc_warp.html}{here}}
\label{fig:3D_plot}
\end{center}
\end{figure*}

% Paragraph#1 (intro)
In this section, we investigate the outer peripheral warp of the LMC—defined here as vertical deviations from the disk plane at radii beyond $\gtrsim 7^{\circ}$—and illustrate its extent and structure using three complementary visualizations (Figures~\ref{fig:Z_Map}, \ref{fig:edge_view}, and \ref{fig:3D_plot}). We begin with a brief review of previous studies of LMC warps and other large-scale structural distortions.

% \begin{figure}
% \includegraphics[width=0.5\textwidth]{Figures/XY_cc_Z_ageCorr_16deg_Do47.6kpc.png}
% \caption{Z-coordinate map in the on-sky cartesian frame, color-coded by the vertical coordinate $Z$ (in kpc). The map is centered on the adopted LMC center ($\alpha$,$\delta$) = ($81.90^{\circ}$,$-69.87^{\circ}$) and distance of 47.6~kpc. Concentric dashed circles indicate angular separations of $7^\circ$ ($\sim$ 6~kpc), $10^\circ$ ($\sim$ 9~kpc), $13^\circ$ ($\sim$ 11.5~kpc), and $16^\circ$ ($\sim$ 14~kpc) from the LMC center. A prominent central spiral feature is visible in red, but rapidly dissipates due to the LMC’s disk inclination. At radii beyond $\sim$7$^\circ$, the peripheral warp becomes strikingly apparent in the vertical structure.} 
% \label{fig:Z_Map}
% \end{figure}

% Paragraph #2
Numerous studies have shown that the LMC departs from a simple flat-disk geometry, as discussed in Section~\ref{sec:inc_Pos}. \citet{Olsen2002} provided some of the earliest evidence for a stellar warp in the southwestern inner disk, using RC stars to measure vertical displacements of up to $\sim$2.5~kpc relative to the best-fit disk plane. This result was later confirmed by \citet{Choi2018a} with deep SMASH photometry \citep{Nidever2017}. More recently, \citet{JimenezArranz2025} used kinematic data for red-giant stars from \textit{Gaia} and SDSS-IV/V to identify substantial warps within the inner $5^{\circ}$ of the disk (see their Figure~13).

% Paragraph #3
The outer peripheral warp of the LMC was first identified in the Dark Energy Survey (DES; \citealt{TheDarkEnergySurveyCollaboration2005}) footprint of the northern disk by \citet{Balbinot2015}, who interpreted it as either a warped and flared in-situ disk or the early signature of a spheroidal halo component. \citet{Choi2018a} later reported evidence for an outer stellar warp in the southern disk, finding a maximum amplitude of $\sim$4~kpc in the southwest at a galactocentric radius of $\sim$9~kpc. More recently, \citet{Saroon2022}, using \textit{Gaia} EDR3 data, detected a corresponding northeastern warp with a smaller amplitude but similar orientation, together suggesting a U-shaped warp.

% \begin{figure*}
% \begin{center}
% \includegraphics[width=1.0\textwidth]{Figures/max_gradient_edge_view_plot_5arcmmin_spatialbin_DistCorr_int_choi.png}
% \caption{Edge-on view of the LMC disk structure.
% \textit{Left:} The 3D distribution of RC stars projected along the axis of maximum line-of-sight depth gradient, plotted against the vertical coordinate $Z$ (in kpc). The inclination of the LMC disk causes the northeastern (NE) side to appear closer and the southwestern (SW) side farther away. The red dashed line shows the best-fit plane to the disk, capturing its global tilt.
% \textit{Right:} The same edge-on projection after correcting for the disk’s inclination. The vertical axis shows the residual distance $\Delta d$ from the best-fit plane. The red curve traces the median residual distance, weighted by the local on-sky stellar density. The yellow curve shows the median profile from \citet{Choi2018a}, over-plotted for comparison.}
% \label{fig:edge_view}
% \end{center}
% \end{figure*}

%Paragraph #4
Simulations by \citet{Besla2012} (see also Figure 15 of \citealt{Choi2018a}) predict a southwest warp extending to larger distances and a northeast warp tilted toward the observer (i.e., closer). Our results, however, together with the northern warps identified by \citet{Balbinot2015} and \citet{Saroon2022}, contradict this picture: both the southwestern and northeastern warps extend away from the observer, suggesting a departure from early theoretical expectations.

%Paragraph #5
U-shaped warps are far less common and more difficult to identify than the more typical S-shaped warps \citep{Ann2006, Zee2022}. They may originate from repeated flyby encounters at varying incidence angles \citep{Kim2014} or from the superposition of asymmetric S-shaped warps \citep{Saha2006}. \citet{Zee2022} also associate U-shaped warps with ram-pressure stripping in jellyfish galaxies, pointing to a potential non-tidal origin.

% Paragraph #6
Figure~\ref{fig:Z_Map} presents the median $Z$–coordinate distribution of LMC stars within $13^{\circ}$ of the center, evaluated in the on-sky Cartesian system. We restrict the analysis to this radius ($\sim$11.4 kpc at 50 kpc) because beyond $13^{\circ}$ the disk begins to exhibit higher-order warp signatures and possible halo-driven distortions, which we defer to a forthcoming study. The adopted origin of the coordinate system, $(\alpha_{0}, \delta_{0}, D_{0})$, is identical to that described in Section~\ref{sec:inc_Pos}.

In this map, red regions indicate stars located closer than $D_{0}$, while blue regions denote stars lying farther away. Black dashed circles mark projected galactocentric radii of $7^{\circ}$ ($\sim$6 kpc), $10^{\circ}$ ($\sim$9 kpc), and $13^{\circ}$ ($\sim$11.4 kpc). We note that this representation differs from the earlier distance map: because the $(X, Y)$ grid is defined in LMC-centric Cartesian coordinates rather than the plane of the sky, it gradually diverges from the true on-sky projection at larger radii. The map reveals a pronounced U-shaped warp in the outer disk, which is nearly azimuthally symmetric, with the periphery systematically displaced to larger distances relative to the inner regions.

% Robustness to coordinate origin
Similar to Section~\ref{sec:inc_Pos}, we apply small perturbations ($\pm \sim 0.50^{\circ}$ or $\pm \sim$0.4~kpc) to the adopted LMC center in the Cartesian coordinate system. We find that the overall qualitative description of the peripheral warp—the global U-shaped displacement and its asymmetric morphology—remains unchanged. However, the precise quantitative values, such as the maximum warp amplitude or the radius at which the warp becomes prominent, shift slightly depending on the adopted center. These tests demonstrate that while the detailed numbers carry some sensitivity to the exact choice of origin, the existence and large-scale structure of the warp are robust features of the data.

% Paragraph #7
The main spiral arm is clearly visible within $7^{\circ}$, appearing as a red feature that gradually transitions to white and then blue due to the disk inclination. Beyond $\sim$7$^{\circ}$, the entire periphery turns blue, indicating that the median $Z$ coordinate lies farther away than $D_{0}$. The DES northern warp reported by \citet{Balbinot2015} covers only a limited portion of the northern periphery, roughly between $X = 0$–5~kpc and $Y = 6$–10~kpc. The southern warp identified by \citet{Choi2018a} appears as a dark blue extension just inside the $10^{\circ}$ circle in the southwest, while the northeastern warp detected by \citet{Saroon2022} appears as a light blue feature between the $7^{\circ}$ and $10^{\circ}$ circles. Our new result is that the entire stellar periphery at all azimuthal angles—at least for intermediate-age stars ($\sim$1.5–10~Gyr)—is displaced outward relative to the inner disk, forming a U-shaped bowl structure. The warp is asymmetric, suggesting that its origin may lie in the superposition of multiple S-shaped warps \citep{Saha2006} or in repeated flyby encounters with the SMC \citep{Kim2014}.

% Figure~\ref{fig:3D_plot} shows the same data as Figure~\ref{fig:Z_Map}, but rendered in a 3D view. An animation of this plot is available online to visualize the LMC warp structure from multiple viewing angles.

%Paragraph #8
An alternate perspective on the LMC’s peripheral warp is shown in Figure~\ref{fig:edge_view}. The left panel displays the 3D distribution of the LMC RC sample, binned into $5^{\prime} \times 5^{\prime}$ spatial cells and projected into the on-sky Cartesian system $(X, Y, Z)$, viewed along the axis of maximum line-of-sight depth (i.e., perpendicular to the line of nodes). The vertical axis corresponds to the $Z$-coordinate, with darker shading indicating higher stellar densities.

The global inclination of the disk is evident: the northeastern side is tilted toward the observer, while the southwestern side lies farther away. The dashed red line marks our best-fit global inclination angle for the inner disk ($i = 29.29^{\circ}$) derived in Section~\ref{sec:inc_Pos}. About this reference plane, the inner disk shows small-scale ripples and localized deviations, consistent with the warps and twists identified by \citet{Olsen2002}, \citet{Choi2018a}, and \citet{JimenezArranz2025}.

%Paragraph #9
The right panel of Figure~\ref{fig:edge_view} presents an edge-on view of the LMC after removing the global disk inclination. In this projection, the vertical axis ($\Delta d$) represents the distance offset relative to our best-fit disk plane (red dashed line). The red curve shows the median $\Delta d$ profile, computed in 0.5~kpc bins within $\pm$5~kpc and in 2~kpc bins at larger radii, with each bin weighted by the on-sky stellar number density. For comparison, the yellow curve plots the corresponding $\Delta d$ profile from \citet{Choi2018a}.

Our measurements closely match the SMASH results within $\pm$5~kpc and along the southwestern disk out to $\sim$9~kpc. The differences that emerge in the northeast are primarily driven by our substantially larger areal coverage compared to SMASH. Moreover, our inferred offsets are consistent with the findings of \citet{Saroon2022}, who reported $\Delta d$ amplitudes of $\sim$1–2~kpc within 8~kpc of the LMC center in the northeast.

% \begin{figure*}
% \begin{center}
% \includegraphics[width=0.75\textwidth]{Figures/3D_XYZ_LMC_colorbar.png}
% \caption{3D plot in the on-sky cartesian coordinate system (X,Y,Z) described in Section~\ref{sec:inc_Pos} of our LMC RC sample out to $\sim$13$^{\circ}$. The colored surface is the median Z-coordinate with blue (red) indicating further (closer) distances than the adopted LMC centric distance of $D_{o}$ = 49.35~kpc. An animation of this plot from different viewing perspectives is available online.}
% \label{fig:3D_plot}
% \end{center}
% \end{figure*}

% Paragraph #10
The dominant warp feature—visible as two prominent ``horns'' extending towards negative $Z$ and $\Delta d$—is strikingly apparent. This peripheral distortion emerges at $\sim\pm$5~kpc ($\sim$$6^{\circ}$) and grows to amplitudes of $\sim$7~kpc by a galactocentric radius of $\sim$15~kpc in the southwest. Notably, both edges of the warp bend toward the southwest, the direction of the SMC and opposite to the LMC’s systemic motion. This geometry strongly hints at a tidal origin, though detailed dynamical modeling is required to confirm the mechanism. In a forthcoming paper, we demonstrate with $N$-body simulations of LMC–SMC interactions that the SMC’s most recent peri-centric passage can induce peripheral ripples that qualitatively reproduce the observed azimuthal warp \citep{Garver2025}.

% Paragraph #11
Another noteworthy result is the behavior of the median $\Delta d$ curve in the northeast. Although this region is significantly lower in stellar density beyond $\sim$10~kpc compared to the southwest, the warp remains coherent, reaching amplitudes of $\sim$3–4~kpc at 15~kpc from the LMC center. This coherence, despite the sparse stellar environment, implies that the warp is not confined to the dense in-situ disk but may also involve the outer, possibly halo-like, stellar populations of the LMC.

% Figure~\ref{fig:3D_plot} shows the same data as Figure~\ref{fig:Z_Map}, but rendered in a 3D view. An animation of this plot is available online to visualize the LMC warp structure from multiple viewing angles.

% Paragraph #12
Finally, we present the LMC peripheral azimuthal warp as a 3D median-$Z$ surface plot in $(X, Y, Z)$ Cartesian coordinates (Figure~\ref{fig:3D_plot}). This visualization displays the same data as Figure~\ref{fig:Z_Map}, but rendered in three dimensions with the viewpoint oriented toward the northwest and slightly above the disk plane for clarity. An accompanying animation is available online\footnote{Link to LMC warp animation can be found \href{https://slateroden.github.io/lmc_warp.html}{here}}, providing multiple viewing angles to better illustrate the full extent of the LMC’s peripheral warp.

% Paragraph #13 (Conclusion)
In summary, our results confirm the inner- and intermediate-radius warp signals previously identified in localized regions by \citet{Olsen2002}, \citet{Balbinot2015}, \citet{Choi2018a}, \citet{Saroon2022}, and \citet{JimenezArranz2025}, while extending the analysis to much larger radii and, crucially, to full azimuthal coverage. With this expanded spatial baseline, we find that the LMC’s intermediate-age stellar periphery is not composed of isolated sectoral distortions but instead forms a continuous, globally coherent warp encircling the galaxy. The periphery is systematically displaced to larger distances relative to the inner disk, producing a U-shaped warp morphology with clear asymmetries. These results demonstrate that the previously reported ``local’’ warp detections are facets of a single, galaxy-scale vertical distortion, providing a new and stringent observational constraint on models of the recent LMC–SMC interaction and the dynamical response of the extended LMC disk and halo.

% In addition to our RC analysis, other stellar tracers such as AGB stars and RR~Lyrae offer the potential for independent assessments of the LMC’s vertical structure. However, both populations require careful treatment of their own age and metallicity dependencies, which is beyond the scope of this work. A detailed multi-tracer comparison will be the subject of future efforts.

% Section #5
\section{Discussion}
\label{sec:discussion}

% Section #5.1
\subsection{Systematic Effects}
\label{sec:systematic_errors}

% Paragraph #1
The primary objective of this work is to characterize the three-dimensional structure, morphology, and distances of the MCs using their intermediate-age RC stellar population. To ensure the robustness of our measurements, we carefully quantify and propagate the dominant sources of systematic uncertainty that could bias our results. In particular, we consider the impact of (i) foreground MW contamination, (ii) residual RGB inclusion in the RC selection, (iii) population-dependent effects, including uncertainties in the RC absolute magnitude calibration and biases introduced by age-dependent variations in RC luminosity, (iv) \textit{Gaia} photometric uncertainties, and (v) uncertainties in the adopted dust extinction corrections.

% Paragraph #2 (need to discuss Besancon model)
% The first source of uncertainty is contamination from MW foreground stars and the inclusion of RGB stars overlapping with the RC locus in the CMD. Even after applying our Magellanic RC selection (Section~\ref{sec:RC_selection}), a small residual level of contamination is unavoidable. For the MW component, halo stars with parallaxes $\varpi \leq 0.15$ mas are expected to occupy the Magellanic PM space at roughly the 1\% level when using a single global PM cut. Our local, field-by-field PM selection reduces this fraction to well below 1\%. If residual MW contamination remains at the $\sim$0.5\% level, the bias introduced can be estimated from the median absolute magnitude offset between MW and MC RC stars in the \textit{Gaia} $G$ band, which we calculate to be $\Delta M_G \simeq 0.295$~mag, using the median $M_G$ from \citet{Ruiz-Dern2018} and our MC RC median $M_G$. This translates to a systematic shift of  $\sim0.0015$~mag, corresponding to a fractional distance error of $\sim$0.07\% ($\sim$0.04~kpc at the distance of the LMC). For the RGB, a simple fixed CMD “RC box” can admit 10–40\% contamination \citep[e.g.,][]{Choi2018a}; however, by explicitly modeling the local red-giant color distribution with a double Gaussian plus second-order polynomial, we constrain the RGB contribution to $<$10\%. Crucially, the subset of RGB stars that pass our local CMD fits have nearly identical colors and magnitudes to genuine RC stars, and therefore do not bias our distance estimates.

% Paragraph #2 (rewritten)
The first source of uncertainty is contamination from MW foreground stars and the inclusion of RGB stars overlapping with the RC locus in the CMD. Even after applying our Magellanic RC selection (Section~\ref{sec:RC_selection}), a small residual level of contamination is unavoidable. To quantify the MW component, we used the Besancon MW model \citep{Robin2003} to generate a synthetic catalog of MW stars over the same square degree coverage as our MC sample ($\sim$2,110 $deg^{2}$) centered on the LMC, applying identical magnitude and color selections to the RC locus in the CMD. We then compared the predicted MW distribution in PM space to that of the Magellanic RC population. From this exercise, we find that RC MW halo stars with parallaxes $\varpi/\sigma_{varpi} \leq 4$ mas are expected to overlap the Magellanic PM space at roughly the 0.1\% level if using a single global $2\sigma_{PM}$ PM cut. Our local, field-by-field PM selection reduces this fraction to well below 0.1\%. Taking an upper limit of MW contamination at the 0.1\% level, the bias introduced can be estimated from the median absolute-magnitude offset between MW and MC RC stars in the \textit{Gaia} $G$ band, which we calculate to be $\Delta M_G \simeq 0.295$~mag, using the median $M_G$ from \citet{Ruiz-Dern2018} and our MC RC median $M_G$. This translates to a systematic uncertainty of $\sim0.0003$~mag, corresponding to a fractional distance error of $\sim$0.01\% ($\sim$0.007~kpc at the distance of the LMC).  

For the RGB, a simple fixed CMD “RC box” can admit 10–40\% contamination \citep[e.g.,][]{Choi2018a}; however, by explicitly modeling the local red-giant color distribution with a double Gaussian plus second-order polynomial, we constrain the RGB contribution to $<$10\%. Crucially, the subset of RGB stars that pass our local CMD fits have nearly identical colors and magnitudes to genuine RC stars, and therefore do not bias our distance estimates.

% Paragraph #3
A second source of systematic uncertainty arises from population effects—specifically the intrinsic dependence of RC luminosity on stellar age and metallicity. As discussed in Sections~\ref{sec:population_systematics} and~\ref{sec:SFH_RC_ages}, the RC absolute magnitude is not universal: younger or more metal-poor populations are systematically brighter, while older or more metal-rich populations are fainter. If these variations are ignored and a single RC absolute magnitude is assumed, RC population distance errors of $\sim$10–15\% can result—equivalent to $\sim$5–7.5~kpc at the distance of the LMC.

% Paragraph #4 (rewritten)
In Section~\ref{sec:abs_mag_calibration}, we calibrated the RC absolute magnitude, $M_G$, as a function of $(G-K_s)$ color. Two terms dominate the associated error budget. The first is the uncertainty in deriving equation~\ref{eq:Mg}, which arises from fitting the modal RC distribution from \citet{Ruiz-Dern2018}. The second comes from converting $(BP-RP)$ color into $(G-K_s)$ color \citep{GaiaCollaboration2021b}. Adding these uncertainties in quadrature yields a total error of $\sim$0.1 mag in $M_G$, corresponding to a distance uncertainty of $\sim$2.3 kpc at the distance of the LMC.

% Paragraph #5 
Uncertainties in the median RC age, and therefore in our second-order RC age–distance bias correction, arise from several factors, including the well-known age–metallicity degeneracy in the CMD, photometric completeness limitations, and the assumptions inherent to the stellar evolutionary models used to construct synthetic CMDs. To quantify these errors, we employed Monte Carlo resampling of the best-fit SFH solutions. For each hybrid bin (see Section~\ref{sec:SFH_RC_ages}), RC stars were first identified within the synthetic CMDs generated by \verb|TheStorm| and mapped to their parent SFH age bins. The star formation rates in each age bin were perturbed according to their fitted uncertainties, with negative draws clipped at zero. Each RC star was then assigned a weight proportional to the perturbed SFR multiplied by the bin width, and a weighted mean RC age was recalculated across all stars. Repeating this procedure 4000 times produced a distribution of mean RC ages, from which we extracted the 16th, 50th, and 84th percentiles. We adopted the central value as the median RC age, with the uncertainty defined as half the difference between the 84th and 16th percentiles. Across all hybrid bins, the mean age uncertainty is $\sim$0.2 Gyr, with a range of $\sim$0.05–1.6 Gyr. Using the \citet{Girardi2001} theoretical RC curves, this corresponds to a median uncertainty in the RC luminosity of $\sim$0.01 mag (or $\sim$0.23 kpc at the distance of the LMC). Adding this RC age–dependent luminosity uncertainty in quadrature with the $M_G$ uncertainty defined above yields a total systematic uncertainty from RC population effects of $\sim$0.10 mag, equivalent to a distance uncertainty of $\sim$2.31 kpc at the distance of the LMC.

%paragraph 6
A further source of error stems from our use of the initial RC distance calculated in Section~\ref{sec:abs_mag_calibration} as a prior in the SFH fitting procedure. While this prior anchors the CMD modeling, it inevitably introduces some sensitivity to the adopted RC distance. However, \verb|TheStorm| explicitly allows for small shifts in the CMD to compensate for mismatches in distance, thereby mitigating the impact of this assumption. As a result, while the prior does contribute to the systematic error budget, its effect is modest compared to the dominant uncertainties associated with RC population effects.

% Our systematic uncertainties Age-related uncertainties stem from the goodness of fit (i.e., reduced chi square value) of the reproduced CMD compared to the observed one, with values on the order of 0.5, corresponding to an uncertainty on the order of $\sim$ 0.5 Gyrs or $\sim$ 0.12 mags. Adding this in quadrature with the 0.15 mag photometric term gives a total population-driven systematic uncertainty of $\sim$ 0.19 mag.

% Paragraph #7
Photometric uncertainties in \textit{Gaia} data also propagate into our RC distance estimates. For the typical apparent magnitude of MC RC stars (G $\sim$ 18.90 mag), the median photometric uncertainty in the Gaia $G$ band is $\sim$0.003~mag. At the distance of the Clouds ($\sim$50–60~kpc), this corresponds to an uncertainty of only $\sim$0.07–0.08~kpc, which is negligible compared to the $\sim$kpc-scale systematic errors associated with population effects. Importantly, the $G$-band magnitude enters directly into the distance modulus, while $(BP-RP)$ colors are only used indirectly through the $M_G$–color calibration. Thus, the $G$-band error acts as a direct systematic on distance, whereas the $BP$ and $RP$ photometric uncertainties propagate indirectly by broadening the color distribution and introducing scatter into the population correction term (see Section~\ref{sec:population_systematics}). Taken together, we estimate that purely photometric contributions amount to $\lesssim$0.01 mag in distance modulus ($\lesssim$0.25 kpc), which is subdominant compared to population-driven systematics but non-negligible in our full error budget.

% Paragraph 8
Finally, uncertainties in our extinction correction, $A_{\lambda}$, propagate into the RC-based distance estimates. These arise from three primary sources: (i) errors in the adopted reddening values $E(B-V)$, (ii) potential deviations from the assumed total-to-selective extinction ratio $R_V$, and (iii) the application of a single extinction law across the Clouds. For the inner regions of the LMC and SMC, we adopt the OGLE-IV reddening map, which provides high spatial resolution but carries a median uncertainty of $\sim$0.04~mag. In the outer, low-density periphery, we rely on the recalibrated SFD maps, where the extinction is typically low and the associated errors negligible. Adopting the \citet{Fitzpatrick1999} extinction law with a fixed $R_V = 3.1$ introduces an additional 1--2\% uncertainty in the dereddened photometry \citep{Schlafly2010}, though in low-extinction regions this is $\ll 1\%$. The largest extinction-related uncertainties arise in dusty star-forming regions such as 30~Doradus in the inner LMC, where patchy dust structures can produce spatial variations not fully captured by the adopted maps. Taken together, we estimate a combined uncertainty in the extinction correction of $\sim$2--3\%, corresponding to a total systematic error of $\sim$0.01~mag, or equivalently $\sim$0.23~kpc at the distance of the LMC.

% With a median extinction value of $A_{G} \simeq 0.18$ mag across our RC sample, 

% Paragraph #9
Combining all identified sources of systematic uncertainty—foreground MW contamination \((\sigma_{\mathrm{MW}}^2)\), RC population effects \((\sigma_{\mathrm{Pop}}^2)\), photometric errors \((\sigma_{\mathrm{Phot}}^2)\), and extinction corrections \((\sigma_{\mathrm{Ext}}^2)\)—together with the random statistical uncertainties \((\sigma_{\mathrm{Stat}}^2)\) (see Section~\ref{sec:MC_maps}), we construct the total error budget as

\begin{equation} 
\label{eq:error}
\sigma_{\mathrm{Total}}^2 = \sigma_{\mathrm{Stat}}^2 + \sigma_{\mathrm{MW}}^2 + \sigma_{\mathrm{Pop}}^2 + \sigma_{\mathrm{Phot}}^2 + \sigma_{\mathrm{Ext}}^2
\end{equation}

\noindent Applying this framework, we derive a revised estimate of the median LMC distance within $\sim$23$^{\circ}$ of its center as \(D_{\mathrm{LMC}} = 50.62 \pm 2.32\)~kpc. For the SMC, within $\sim$12$^{\circ}$, we obtain \(D_{\mathrm{SMC}} = 60.75 \pm 2.85\)~kpc. These values represent the most spatially extended median distance determinations for the Clouds to date, encompassing not only their central regions but also their diffuse, intermediate-age stellar peripheries.

% Paragraph #10
Our final median distances for both the LMC and SMC are fully consistent with previous high-precision determinations. For the LMC, our value of $D_{\mathrm{LMC}} = 50.62 \pm 2.32$~kpc agrees closely with the eclipsing binary distance of $49.59 \pm 0.54$~kpc from \citet{Pietrzynski2019} and with other independent distance measurements \citep[e.g.,][]{vandermarel2001b, deGrijs2015, Choi2018a}. For the SMC, our result of $D_{\mathrm{SMC}} = 60.75 \pm 2.85$~kpc is within $\sim$1.7~kpc of the eclipsing binary distance of $62.44 \pm 0.81$~kpc from \citet{Graczyk2020}, again in good agreement with the literature and comfortably within our adopted error budget. This consistency underscores the robustness of our methodology across different data sets and approaches. We note, however, that additional sources of error not explicitly quantified here—such as uncertainties in stellar evolutionary models and in the treatment of distance priors within our SFH fitting—are likely still present at some level. These factors may shift the absolute scale slightly, but the close agreement with independent benchmarks gives confidence that our distance framework is both reliable and broadly applicable.

% To correct our LOS depth estimates of the Clouds for systematics, we can follow a similar approach as \citet{Subramanian2009} by subtracting out the dispersion in the intrinsic RC population effects and the observational error in the color from the observed dispersion in the magnitude to derive the dispersion due solely to the depth of the clouds.
% Therefore, subtracting in quadrature the dispersion in color from the dispersion in magnitude, we get a geometrical dispersion for the inner $7^{\circ}$ of the LMC of 0.146 mag, which at a median distance of 50.69 corresponds to a FWHM line-of-sight thickness of $\sim$ 3.5 kpc, which is in excellent agreement with the value of 3.44 ± 1.4 kpc reported by \citet{Subramanian2009}. Applying the same technique, we get an average line-of-sight depth for the outer LMC ($>$ $7^{\circ}$) of $\sim$ 5.2 kpc. For the SMC we obtain an average depth of $\sim$ 6.8 kpc. These systematic corrected depth estimates of the clouds are in much better agreement with previous literature estimates \citealp[]{Subramanian2009,,JacyszynDobrzeniecka2017,YanchulovaMericaJones2017,Choi2018a}.

% Paragraph #11
Furthermore, our detection of the azimuthal peripheral warp of the LMC is robust against all identified sources of systematic uncertainty. Errors from extinction correction, photometry, MW contamination, and RC population effects are far too small to reproduce the large-scale vertical deviations observed in the outskirts. Among these, RC population effects are the dominant contributor, yet even in the most extreme case they account for at most $\sim$50\% of the observed warp amplitude prior to correction. After explicitly correcting for population effects—following the procedures in Sections~\ref{sec:population_systematics} and~\ref{sec:SFH_RC_ages}—the warp remains strong, coherent, and azimuthally continuous. We also tested the sensitivity of our results to modest shifts in the adopted LMC center and fiducial distance ($D_{0}$); while such perturbations slightly change the measured amplitude, the overall morphology and its full azimuthal extent remain unaffected. Taken together with the independent detections of similar distortions in previous studies \citep{Balbinot2015,Choi2018a,Saroon2022}, these tests demonstrate that the peripheral warp cannot be explained as an artifact of measurement or modeling assumptions. Instead, it must represent a genuine geometric distortion in the intermediate-age stellar periphery of the LMC.

% % Section #5.2
% \subsection{Comparison with Independent Distance Tracers}
% \label{sec:independent_tracers}

% While this study relies on RC stars as standard candles, other stellar populations offer independent distance estimates to the Magellanic Clouds and may provide valuable cross-checks on our results. In particular, AGB stars have been used to 
% trace large-scale structures in the LMC and SMC, and modal magnitude techniques 
% can in principle constrain line-of-sight distance variations. However, their 
% limited depth and sensitivity at large radii prevent a robust verification of the 
% peripheral warp identified in this work. RR~Lyrae stars, by contrast, extend 
% throughout the Magellanic periphery, but their pulsational properties and orbital 
% distributions complicate direct comparisons with the intermediate-age RC population. 
% Indeed, we do not detect the same pronounced warp signal in the RR~Lyrae population, 
% which may reflect their different dynamical histories rather than an inconsistency 
% in distance determination. A detailed multi-tracer analysis, accounting for the 
% distinct age and metallicity dependencies of each population, lies beyond the scope 
% of this paper but will be an important avenue for future work.

% Section #5.3
\subsection{Comparison with Independent Distance Tracers}
\label{sec:independent_distance_tracers}

%Discuss brief comparison with independent distance tracers (e.g., TRGB, AGB Modal magnitudes, RR Lyrae stars).

While this study relies on RC stars as standard candles, several other stellar populations provide independent distance constraints to the MCs and therefore offer valuable external checks on our results. Established tracers—including AGB modal magnitudes, TRGB magnitudes, and pulsating variables such as RR Lyrae and Cepheids—have long yielded accurate and mutually consistent distance estimates across both Clouds \citep{vandermarel2001a,Ripepi2015,Pietrzynski2019,Graczyk2020}. 

For example, \citet{Madore2025} recently used calibrated AGB modal magnitudes to derive distance moduli of 18.48 mag ($\sim$49.7 kpc) for the LMC and 18.98 mag ($\sim$62.5 kpc) for the SMC. TRGB-based measurements from \citet{Groenewegen2019} similarly produce distance moduli of $\sim$18.48–18.57 mag ($\sim$49.7–51.7 kpc) for the LMC and $\sim$18.92–19.07 mag ($\sim$60.8–65.1 kpc) for the SMC. Variable-star distances show the same pattern: \citet{Borissova2009} reported an LMC distance modulus of 18.53 mag ($\sim$50.8 kpc) from RR Lyrae stars, while \citet{Haschke2012a} found 18.62 mag ($\sim$53.1 kpc) from RR Lyrae and 18.65 mag ($\sim$53.9 kpc) from Cepheids. In the SMC, \citet{Haschke2012b} derived distance moduli of 18.94 mag ($\sim$61.5 kpc) from RR Lyrae and 19.00 mag ($\sim$63.1 kpc) from Cepheids.
While these methods are robust and yield highly accurate and mutually consistent distance estimates, they suffer from intrinsically low number statistics—particularly in the diffuse, low–surface–brightness periphery of the Clouds. As a result, very few studies have explored distances to specific peripheral substructures using these independent tracers. Yet, confirming the LMC’s peripheral warp with such methods would provide the strongest possible validation of our results. To our knowledge, however, no independent measurement of periphery distances using AGB stars, TRGB stars, RR Lyrae, or Cepheids currently exists for direct comparison.

As an initial attempt to address this gap, we queried \textit{Gaia} data for AGB candidates and estimated distances using AGB magnitudes. We applied selection criteria consistent with those used throughout this work to minimize MW contamination and isolate genuine Magellanic AGB stars. Candidates were identified using a slanted selection box that traces the AGB sequence in the extinction-corrected CMD. After these cuts, only $\sim$33,000 LMC AGB stars remained, extending to galactocentric radii of roughly $8^{\circ}$—just far enough to begin probing the onset of the peripheral warp.

The resulting distance estimates show encouraging behavior: in the inner periphery, the inferred distances begin to shift outward in a manner broadly consistent with the expected warp signature. Because AGB stars span similar age and metallicity regimes across the Clouds, their distance distributions should be closely comparable; thus even modest trends in their inferred distances can provide useful, independent support for the warp geometry.

We also compared our RC-based distances with those derived from a carefully selected \textit{Gaia} EDR3 RR Lyrae sample in the Clouds. This RR Lyrae sample yields an overall median LMC distance of $\sim$51.5 kpc, consistent with previous determinations; however, unlike the RC population, they do not exhibit a pronounced U-shaped warp signature in the periphery. This discrepancy is not entirely unexpected: RR Lyrae are ancient, metal-poor stars that predominantly trace the LMC’s old, roughly spheroidal halo rather than its disk. Consequently, their line-of-sight distance distribution need not mirror that of the intermediate-age RC population, which resides primarily in the disk where the warp is imprinted. Nevertheless, the contrasting peripheral distance trends between these two independent tracers are intriguing and warrant further investigation in future studies.

Taken together, these exploratory comparisons highlight both the promise and the current limitations of using independent stellar tracers to validate the LMC’s peripheral warp. The sparse sampling of AGB stars and the fundamentally different spatial distribution of RR Lyrae stars limit their ability to fully trace the warped disk, yet the preliminary signals we do detect—particularly from the AGB population—suggest that complementary distance indicators may ultimately corroborate at least part of the RC-based warp signature. A definitive multi-tracer confirmation will require deeper, wider, and more uniformly sampled data, ideally combining future \textit{Gaia} releases with sensitive near-IR surveys capable of capturing AGB, TRGB, and variable stars across the extreme outskirts of the Clouds. Such efforts will be essential for establishing an independent, population-agnostic view of the Magellanic periphery and for placing the geometric warp on firmer empirical footing.

% In addition, the azimuthal peripheral warp of the LMC has remained elusive in these independent distance tracers. This is largely in part due to the limited coverage and low number statistics of these methods (e.g., Cepheids, TRGB and AGB modal magnitudes). Nothing would solidify the fact of the LMC warp more than finding similar distance results from these independent distance tracers. 

% Section #5.3
\subsection{Constraints on LMC-SMC Orbital History}
\label{sec:orbital_history}

% Paragraph #1
Although independent distance tracers are not yet numerous enough to map the extreme Magellanic periphery, their preliminary consistency with our RC-based distances supports the large-scale geometric features identified here—including the fully azimuthal warp of the LMC. Combined with the systematic-error analysis in Section~\ref{sec:systematic_errors}, these cross-checks provide a secure empirical basis for interpreting the three-dimensional structure of the system. This framework is crucial for assessing the dynamical origin of the warp, since any viable model of the recent LMC–SMC interaction must reproduce the absolute distance scale, the coherent vertical distortions, and the radial changes in disk geometry that we measure. We therefore turn to the implications of these structural constraints for the orbital history of the Clouds, focusing particularly on what the far-periphery distances reveal about the timing, geometry, and impact of their most recent encounters.

% Paragraph #1
Our detailed mapping of the Magellanic morphology and structure provides stringent new constraints on the recent orbital history of the Clouds. By measuring individual distances across the full stellar extents of both the LMC and SMC, we obtain the most spatially extended distance determinations to date, establishing the present-day three-dimensional configuration of their intermediate-age populations. Modern simulations are now capable of reproducing peripheral features such as the LMC’s Northern Arm and the Magellanic Bridge, and our distance measurements to these structures supply critical benchmarks for testing and refining these models. In particular, the precisely mapped locations of the Northern Arm, Magellanic Bridge, and other outer-disk substructures serve as powerful boundary conditions for $N$-body modeling, directly informing efforts to reconstruct their origin and subsequent dynamical evolution.

% Paragraph #2
Our measurements of the LMC disk geometry reveal substantial departures from a simple, flat configuration—consistent with earlier studies, but now traced to far larger radii. Within the inner disk ($R < 7^{\circ}$), our derived inclination and line-of-nodes position angle agree well with previous determinations. Beyond this radius, however, the geometry changes dramatically: the inclination declines steadily, while the position angle first decreases and then rises sharply. Together, these trends indicate a morphologically and kinematically disturbed outer disk, plausibly shaped by tidal torques from past interactions with the SMC and/or the MW. We encourage future simulations to track the evolution of $(i,\theta)$ as a function of time and galactocentric radius, as our present-day constraints at large projected distances provide powerful leverage for reconstructing the recent dynamical history of the Magellanic system.  

% Paragraph #3
Our distance measurements in the far periphery of the LMC reveal a fully azimuthal warp extending to radii of $\sim$13$^{\circ}$ and reaching amplitudes of up to $\sim$7~kpc. This pronounced, U-shaped vertical distortion provides new constraints on the present-day structure of the LMC’s outskirts and offers crucial insight into the timing and nature of its recent interactions. We interpret the warp as the cumulative imprint of repeated close passages of the SMC. Supporting this scenario, \citet{Garver2025} present $N$-body simulations showing that only two peri-centric encounters are sufficient to generate a warp with the observed amplitude and morphology. Their models further predict vertical oscillations in the LMC periphery following the most recent LMC–SMC collision $\sim$150~Myr ago—producing ripple-like patterns strikingly similar to those detected in our data.

% Paragraph #4
In a forthcoming paper \citep{Oden2026a}, we will build directly on the results presented here and by \citet{Garver2025} by quantitatively comparing our observational warp constraints with tailored $N$-body simulations of the LMC–SMC interaction. This analysis will allow us to test dynamical models of the Magellanic System with unprecedented precision, placing tight constraints on the timing of the most recent encounter, the SMC’s impact parameter, and the incident angle of its collision with the LMC. These parameters serve as essential boundary conditions for reconstructing the recent orbital evolution of the Clouds.

% SMC NE closer than LMC SW periphery 

% discuss N LMC dist dispersion higher than S LMC disk

% Inc of inner LMC versus outer LMC

% LMC warp in South and North

% NE SMC large distance dispersion

% NA and LMC hook

% SMC NOD distance and dispersion

% Section #6
\section{Summary}
\label{sec:conclusion}

% Paragraph #1
In this study, we presented a comprehensive analysis of the three-dimensional structure and morphology of the Magellanic Clouds (MCs) using a carefully selected sample of $\sim$2.3 million RC stars from \textit{Gaia} DR3. We demonstrated that \textit{Gaia} is intrinsically complete at the RC magnitude and that crowding-driven incompleteness in the LMC bar and central SMC has a negligible impact on our RC selection. To obtain accurate distances, we implemented a population-corrected RC methodology that accounts for spatial variations in metallicity and stellar age across both Clouds. This approach combines the empirically calibrated color–magnitude relation of \citet{Ruiz-Dern2018} with spatially resolved SFHs derived from deep DELVE–MC photometry \citep{Oden2026b}. Applying this framework, we measured median distances out to $\sim$23$^{\circ}$ for the LMC and $\sim$12$^{\circ}$ for the SMC, yielding $D_{\mathrm{LMC}} = 50.62 \pm 2.32$~kpc and $D_{\mathrm{SMC}} = 60.75 \pm 2.85$~kpc. These represent the most accurate and spatially extended median distances yet obtained for both systems..  

% Paragraph #2
These distances enabled us to construct a high-resolution, contiguous map of the median RC distance across the full Magellanic stellar system. This map reveals not only the inclined LMC disk but also the three-dimensional morphology of prominent peripheral substructures, including the LMC Northern Arm, the two Southern Hooks, the SMCNOD, and the SMCSOD. Most importantly, it provides the first unambiguous evidence for a fully azimuthal warp in the outskirts of the LMC. Complementary maps of the vertical $Z$-coordinate in an on-sky Cartesian frame, together with edge-on projections, emphasize the vertical structure of both the disk and its periphery. To aid visualization and interpretation, we also present a 3D animation of the LMC’s median $Z$-surface, allowing its geometry to be explored from multiple viewing angles. 

% Paragraph #3
A central finding of this work is the discovery of an extended, azimuthally continuous warp in the far periphery of the LMC. This feature emerges beyond $\sim$7$^{\circ}$ and reaches vertical amplitudes of up to $\sim$7~kpc at galactocentric radii of $\sim$15~kpc. The warp exhibits a striking, asymmetric U-shaped geometry, with both the southwestern and northeastern outskirts bending away from the observer—contrary to earlier simulations that predicted opposing signs in these regions. We interpret this structure as a large-scale tidal response of the outer LMC disk, likely imprinted by a recent close encounter with the SMC.  

% Paragraph #4
Taken together, these results provide stringent new constraints on the recent orbital history of the MCs. The detailed distances presented here offers powerful boundary conditions for dynamical models seeking to reconstruct their past interactions and morphological evolution. Forthcoming spectroscopic data from the SDSS-V Magellanic Genesis Survey \citep[MGS;][]{Nidever2025b} and the 4-meter Multi-Object Spectroscopic Telescope \citep[4MOST;][]{deJong2019} will be particularly valuable for probing the outer peripheries of both Clouds. By providing kinematic and chemical information in these diffuse, low–surface–brightness regions, MGS and 4MOST will enable direct tests of our structural measurements and will play a crucial role in mapping the dynamical nature of the LMC’s peripheral warp.

% Write Acknowledgments
\section*{Acknowledgments}

S.J.O. and D.L.N. acknowledge support from National Science Foundation grants AST 1908331 and 2408159.

%We would like to thank the anonymous referee for useful comments that improved the manuscript.

This work has made use of data from the European Space Agency (ESA) mission
{\it Gaia} (\url{https://www.cosmos.esa.int/gaia}), processed by the {\it Gaia}
Data Processing and Analysis Consortium (DPAC,
\url{https://www.cosmos.esa.int/web/gaia/dpac/consortium}). Funding for the DPAC
has been provided by national institutions, in particular the institutions
participating in the {\it Gaia} Multilateral Agreement.

This work as also made use of data obtained with the Dark Energy Camera (DECam) at the Blanco 4m telescope at Cerro Tololo Inter-American Observatory. DECam was constructed by the Dark Energy Survey (DES) collaborating institutions: Argonne National Lab, University of California Santa Cruz, University of Cambridge, Centro de Investigaciones Energeticas, Medioambientales y Tecnologicas-Madrid, University of Chicago, University College London, DES-Brazil consortium, University of Edinburgh, ETH-Zurich, University of Illinois at Urbana-Champaign, Institut de Ciencies de l'Espai, Institut de Fisica d'Altes Energies, Lawrence Berkeley National Lab, Ludwig-Maximilians Universit\"at, University of Michigan, National Optical Astronomy Observatory, University of Nottingham, Ohio State University, University of Pennsylvania, University of Portsmouth, SLAC National Lab, Stanford University, University of Sussex, and Texas A\&M University. Funding for DES, including DECam, has been provided by the U.S. Department of Energy, National Science Foundation, Ministry of Education and Science (Spain), Science and Technology Facilities Council (UK), Higher Education Funding Council (England), National Center for Supercomputing Applications, Kavli Institute for Cosmological Physics, Financiadora de Estudos e Projetos, Funda\c{c}\~ao Carlos Chagas Filho de Amparo a Pesquisa, Conselho Nacional de Desenvolvimento Cientfico e Tecnol\'ogico and the Minist\'erio da Ci\^encia e Tecnologia (Brazil), the German Research Foundation-sponsored cluster of excellence "Origin and Structure of the Universe" and the DES collaborating institutions. The Cerro Tololo Inter-American Observatory, National Optical Astronomy Observatory is operated by the Association of Universities for Research in Astronomy (AURA) under a cooperative agreement with the National Science Foundation. 

The DELVE Collaboration gratefully acknowledges support from
Fermilab LDRD L2019-011, the NASA Fermi Guest Investigator Program Cycle 9 No.
91201, and the National Science Foundation under Grant No. AST-2108168,
AST-2108169, AST-2307126, and AST-2407526. This research is partially funded by a
generous gift from Charles Simonyi to the NSF Division of Astronomical Sciences. The
award is made in recognition of significant contributions to Rubin Observatory’s Legacy
Survey of Space and Time

% \clearpage
% \twocolumn 

\appendix

\section{\textit{Gaia} DR3 Archive Query}

The data for this study were acquired through the ESA \textit{Gaia} DR3 archive. We first queried a large $30^{\circ}$ aperture covering both Clouds with the following ADQL query:

\begin{lstlisting}[style=SQLStyle]
SELECT * 
FROM gaiadr3.gaia_source as g
WHERE 1 = CONTAINS(POINT('ICRS', g.ra, g.dec), 
                   CIRCLE('ICRS', 81.28, -69.78, 30))
  AND g.phot_g_mean_mag BETWEEN 17 AND 20.25
  AND g.bp_rp BETWEEN 0 AND 4.75
  AND g.astrometric_excess_noise < 2.5
  AND g.ruwe < 1.4
  AND g.parallax IS NOT NULL;
\end{lstlisting}

This query slightly truncated the far north-eastern SMC periphery. Therefore, we additionally applied a second, smaller $15^{\circ}$ query centered on the SMC to ensure adequate sampling of the outer SMC periphery:

\begin{lstlisting}[style=SQLStyle]
SELECT * 
FROM gaiadr3.gaia_source as g
WHERE 1 = CONTAINS(POINT('ICRS', g.ra, g.dec), 
                   CIRCLE('ICRS', 13.30, -72.85, 15))
  AND g.phot_g_mean_mag BETWEEN 17 AND 20.25
  AND g.bp_rp BETWEEN 0 AND 4.75
  AND g.astrometric_excess_noise < 2.5
  AND g.ruwe < 1.4
  AND g.parallax IS NOT NULL;
\end{lstlisting}

These catalogs were then combined, retaining only unique objects, to form the main \textit{Gaia} giants sample from which we subsequently obtain our RC subsample.

%\begin{bibliography} %asstex does this already

% \bibliographystyle{aasjournal}
% \bibliography{ref} 

%\end{bibliography}

\end{document}